\documentclass[
twocolumn,
]{aastex631}

\usepackage{amsmath}
\usepackage{graphicx}
\usepackage{xcolor}
\usepackage{xparse}
\usepackage{bm}
\usepackage{latexsym}
\usepackage[
capitalize,
noabbrev
]{cleveref}
\usepackage{tikz}
\usepackage[normalem]{ulem}
\usepackage{paralist}
\usepackage{multirow}

\newcommand{\Bb}{\ensuremath{\bm{b}}}

\newcommand{\Bv}{\ensuremath{\bm{v}}}

\newcommand{\Bz}{\ensuremath{\bm{z}}}

\newcommand{\BB}{\ensuremath{\bm{B}}}

\NewDocumentCommand{\curl}{o o}{\ensuremath{\IfNoValueTF{#2}{\nabla}{\nabla_{#2}} \times #1}}
\NewDocumentCommand{\grad}{o o}{\ensuremath{\IfNoValueTF{#2}{\nabla}{\nabla_{#2}} #1}}
\NewDocumentCommand{\diverg}{o o}{\ensuremath{\IfNoValueTF{#2}{\nabla}{\nabla_{#2}} \cdot #1}}

\NewDocumentCommand{\years}{o}{\ensuremath{
\IfNoValueF{#1}{#1 \,}
\mathrm{years}
}}

\NewDocumentCommand{\days}{o}{\ensuremath{
\IfNoValueF{#1}{#1 \,}
\mathrm{days}
}}

\NewDocumentCommand{\months}{o}{\ensuremath{
\IfNoValueF{#1}{#1 \,}
\mathrm{months}
}}

\NewDocumentCommand{\s}{o}{\ensuremath{
\IfNoValueF{#1}{#1 \,}
\mathrm{s}
}}

\NewDocumentCommand{\nT}{o}{\ensuremath{
\IfNoValueF{#1}{#1 \,}
\mathrm{nT}
}}

\NewDocumentCommand{\km}{o}{\ensuremath{
\IfNoValueF{#1}{#1 \,}
\mathrm{km}
}}

\NewDocumentCommand{\au}{o}{\ensuremath{
\IfNoValueF{#1}{#1 \,}
\mathrm{AU}
}}

\NewDocumentCommand{\amu}{o}{\ensuremath{
\IfNoValueF{#1}{#1 \,}
\mathrm{amu}
}}

\NewDocumentCommand{\temp}{o}{\ensuremath{
\IfNoValueF{#1}{#1 \times}
\mathrm{10^5 \; K}
}}

\NewDocumentCommand{\pct}{o}{\ensuremath{
\IfNoValueF{#1}{#1 \;}
\%
}}

\NewDocumentCommand{\Rs}{o}{\ensuremath{
\IfNoValueF{#1}{#1 \;}
\mathrm{R_S}
}}

\NewDocumentCommand{\kms}{o}{\ensuremath{
\IfNoValueF{#1}{#1 \;}
\mathrm{km \, s^{-1}}
}}

\NewDocumentCommand{\cc}{o}{\ensuremath{
\IfNoValueF{#1}{#1 \;}
\mathrm{cm^{-3}}
}}

\NewDocumentCommand{\mWcc}{o}{\ensuremath{
\IfNoValueF{#1}{#1 \;}
\mathrm{mW \cc}
}}

\NewDocumentCommand{\mWmsq}{o}{\ensuremath{
\IfNoValueF{#1}{#1 \;}
\mathrm{mW \, m^{-2}}
}}

\NewDocumentCommand{\eV}{o}{\ensuremath{
\IfNoValueF{#1}{#1 \;}
\mathrm{eV}
}}

\NewDocumentCommand{\keV}{o}{\ensuremath{
\IfNoValueF{#1}{#1 \;}
\mathrm{keV}
}}

\NewDocumentCommand{\MeV}{o}{\ensuremath{
\IfNoValueF{#1}{#1 \;}
\mathrm{MeV}
}}

\NewDocumentCommand{\nucleon}{s o}{\ensuremath{
\IfNoValueF{#2}{#2 \;}
\IfBooleanTF{#1}{\mathrm{nucleon}}{\mathrm{nuc}}
}}

\NewDocumentCommand{\MeVnuc}{s o}{\ensuremath{
\IfNoValueF{#2}{#2 \;}
\MeV \! /\IfBooleanTF{#1}{\nucleon*}{\nucleon}
}}

\NewDocumentCommand{\keVe}{o}{\ensuremath{
\IfNoValueF{#1}{#1 \;}
\mathrm{keV/e}
}}

\NewDocumentCommand{\Element}{m}{\ensuremath{\mathrm{#1}}}
\NewDocumentCommand{\QState}{m m}{\ensuremath{\mathrm{#1}^{#2+}}}

\NewDocumentCommand{\Hy}{o}{\IfNoValueTF{#1}{\Element{H}}{\QState{H}{#1}}}
\NewDocumentCommand{\He}{o}{\IfNoValueTF{#1}{\Element{He}}{\QState{He}{#1}}}
\NewDocumentCommand{\C}{o}{\IfNoValueTF{#1}{\Element{C}}{\QState{C}{#1}}}
\NewDocumentCommand{\N}{o}{\IfNoValueTF{#1}{\Element{N}}{\QState{N}{#1}}}
\NewDocumentCommand{\Ox}{o}{\IfNoValueTF{#1}{\Element{O}}{\QState{O}{#1}}}
\NewDocumentCommand{\Ne}{o}{\IfNoValueTF{#1}{\Element{Ne}}{\QState{Ne}{#1}}}
\NewDocumentCommand{\Mg}{o}{\IfNoValueTF{#1}{\Element{Mg}}{\QState{Mg}{#1}}}
\NewDocumentCommand{\Si}{o}{\IfNoValueTF{#1}{\Element{Si}}{\QState{Si}{#1}}}
\NewDocumentCommand{\Su}{o}{\IfNoValueTF{#1}{\Element{S}}{\QState{S}{#1}}}
\NewDocumentCommand{\Ca}{o}{\IfNoValueTF{#1}{\Element{Ca}}{\QState{Ca}{#1}}}
\NewDocumentCommand{\Fe}{o}{\IfNoValueTF{#1}{\Element{Fe}}{\QState{Fe}{#1}}}

\NewDocumentCommand{\FIP}{s o O{=}}{\ensuremath{\mathrm{FIP}
\IfNoValueF{#2}{
#3
\IfBooleanTF{#1}{#2}{\eV[#2]}}
}}

\NewDocumentCommand{\AbSEP}{O{X} O{\Ox}}{\ensuremath{#1/#2}}

\NewDocumentCommand{\PLawExp}{s o}{\ensuremath{b
\IfNoValueF{#2}{\IfBooleanTF{#1}{\approx}{=} #2}}}

\newcommand{\he}{\Element{He}}

\NewDocumentCommand{\MpQ}{o}{\ensuremath{
\IfNoValueTF{#1}{\mathrm{M/Q}}{(\mathrm{M/Q})_{#1}}}}

\NewDocumentCommand{\n}{o o O{=}}{\ensuremath{n
\IfNoValueF{#1}{_{#1}}
\IfNoValueF{#2}{
\IfNoValueTF{#3}{=}{#3} \cc[#2]
}
}}

\NewDocumentCommand{\dn}{o o O{=}}{\ensuremath{\delta n
\IfNoValueF{#1}{_{#1}}
\IfNoValueF{#2}{
\IfNoValueTF{#3}{=}{#3} \cc[#2]
}
}}

\NewDocumentCommand{\dnn}{s o o O{=}}{\ensuremath{
\IfBooleanTF{#1}{\dn[#2] / \n[#2]}{\abs{\dn[#2] / \n[#2]}}
\IfNoValueF{#3}{
\IfNoValueTF{#4}{=}{#4} {#3}
}
}}

\NewDocumentCommand{\Wk}{o o O{=}}{\ensuremath{W
\IfNoValueTF{#1}{_k}{_{k,#1}}
\IfNoValueF{#2}{
\IfNoValueTF{#3}{=}{#3} \mWmsq[#2]
}
}}

\NewDocumentCommand{\vsw}{s o O{=}}{\ensuremath{v_\sw
\IfNoValueF{#2}{
\IfNoValueTF{#3}{=}{#3}
\IfBooleanTF{#1}{#2}{\kms[#2]}}
}}

\NewDocumentCommand{\vs}{s o O{=}}{\ensuremath{v_s
\IfNoValueF{#2}{
\IfNoValueTF{#3}{=}{#3}
\IfBooleanTF{#1}{#2}{\kms[#2]}}
}}

\NewDocumentCommand{\vslow}{s o O{=}}{\ensuremath{v_\mathrm{slow}
\IfNoValueF{#2}{
\IfNoValueTF{#3}{=}{#3}
\IfBooleanTF{#1}{#2}{\kms[#2]}}
}}

\NewDocumentCommand{\vfast}{s o O{=}}{\ensuremath{v_\mathrm{fast}
\IfNoValueF{#2}{
\IfNoValueTF{#3}{=}{#3}
\IfBooleanTF{#1}{#2}{\kms[#2]}}
}}

\NewDocumentCommand{\valpha}{s o O{=}}{\ensuremath{v_\alpha
\IfNoValueF{#2}{
\IfNoValueTF{#3}{=}{#3}
\IfBooleanTF{#1}{#2}{\kms[#2]}}
}}

\NewDocumentCommand{\vau}{s o O{=}}{\ensuremath{v_\mathrm{E}
\IfNoValueF{#2}{
\IfNoValueTF{#3}{=}{#3}
\IfBooleanTF{#1}{#2}{\kms[#2]}}
}}

\NewDocumentCommand{\vWk}{s o O{=}}{\ensuremath{v_W
\IfNoValueF{#2}{
\IfNoValueTF{#3}{=}{#3}
\IfBooleanTF{#1}{#2}{\kms[#2]}}
}}

\NewDocumentCommand{\vi}{s o O{=}}{\ensuremath{v_i
\IfNoValueF{#2}{
\IfNoValueTF{#3}{=}{#3}
\IfBooleanTF{#1}{#2}{\kms[#2]}}
}}

\NewDocumentCommand{\vv}{s o O{=}}{\ensuremath{v_v
\IfNoValueF{#2}{
\IfNoValueTF{#3}{=}{#3}
\IfBooleanTF{#1}{#2}{\kms[#2]}}
}}

\NewDocumentCommand{\vn}{s o O{=}}{\ensuremath{v_n
\IfNoValueF{#2}{
\IfNoValueTF{#3}{=}{#3}
\IfBooleanTF{#1}{#2}{\kms[#2]}}
}}

\NewDocumentCommand{\As}{s o O{=}}{\ensuremath{A_s
\IfNoValueF{#2}{
\IfNoValueTF{#3}{=}{#3}
\IfBooleanTF{#1}{#2}{#2 \%}}
}}

\NewDocumentCommand{\rc}{d<> o O{=}}{\ensuremath{r_{c\IfNoValueF{#1}{;{#1}}}
\IfNoValueF{#2}{#3 \Rs[#2]}
}}

\NewDocumentCommand{\rA}{d<> o O{=}}{\ensuremath{r_{A\IfNoValueF{#1}{;{#1}}}
\IfNoValueF{#2}{#3 \Rs[#2]}
}}

\NewDocumentCommand{\grate}{o o}{\ensuremath{
\gamma\IfNoValueF{#1}{/\Omega_{#1}}
\IfNoValueF{#2}{= 10^{{#2}}}
}}

\NewDocumentCommand{\gmax}{o}{\ensuremath{
\gamma_\mathrm{max}\IfNoValueF{#1}{/\Omega_{#1}}
}}

\NewDocumentCommand{\kvec}{o}{\ensuremath{
\vec{k} \rho\IfNoValueF{#1}{{_{#1}}}
}}

\NewDocumentCommand{\kpar}{o}{\ensuremath{
{k_\parallel} \rho\IfNoValueF{#1}{{_{#1}}}
}}

\NewDocumentCommand{\kper}{o}{\ensuremath{
{k_\perp} \rho\IfNoValueF{#1}{{_{#1}}}
}}

\NewDocumentCommand{\ani}{s o}{\ensuremath{
R\IfNoValueF{#2}{_{#2}}
\IfBooleanT{#1}{\, [\perp\!/\!\parallel]}
}}

\NewDocumentCommand{\Temp}{o}{\ensuremath{T{\IfNoValueF{#1}{_{#1}}}}}

\NewDocumentCommand{\Trat}{s m m o}{\ensuremath{
T_{\IfNoValueF{#4}{{#4};}#2}/T_{\IfNoValueF{#4}{{#4};}#3}
 \IfBooleanT{#1}{\, [\#]}
}}

\NewDocumentCommand{\pbeta}{s o}{\ensuremath{
\beta\IfNoValueF{#2}{_{#2}}
 \IfBooleanT{#1}{\, [\#]}
}}

\NewDocumentCommand{\pbetaR}{o}{\ensuremath{
(\pbeta[\parallel
\IfNoValueF{#1}{;#1}], \ani[#1])
}}

\NewDocumentCommand{\dv}{o}{\ensuremath{\Delta v\IfNoValueF{#1}{_{#1}}}}
\NewDocumentCommand{\ca}{o}{\ensuremath{C_{A\IfNoValueF{#1}{;#1}}}}
\NewDocumentCommand{\dvca}{o o}{\ensuremath{\dv[#1]/\ca[#2]}}

\NewDocumentCommand{\nuc}{o}{\ensuremath{\nu_{c\IfNoValueF{#1}{;#1}}}}
\NewDocumentCommand{\Nc}{o}{\ensuremath{N_{c\IfNoValueF{#1}{;#1}}}}
\NewDocumentCommand{\Ac}{o}{\ensuremath{A_{c\IfNoValueF{#1}{;#1}}}}

\NewDocumentCommand{\tauEXP}{o}{\ensuremath{
\tau_{\mathrm{exp}\IfNoValueF{#1}{;#1}
}}}

\NewDocumentCommand{\tauCC}{o}{\ensuremath{
\tau_{\mathrm{C}\IfNoValueF{#1}{;#1}
}}}

\NewDocumentCommand{\SSN}{o}{\ensuremath{\mathrm{SSN}
\IfNoValueF{#1}{#1}}}
\NewDocumentCommand{\NSSN}{o}{\ensuremath{\mathrm{NSSN}
\IfNoValueF{#1}{#1}}}

\newcommand{\sw}{\ensuremath{\mathrm{sw}}}

\NewDocumentCommand{\qpar}{o}{\ensuremath{
q_{\parallel
\IfNoValueF{#1}{;#1}
}}}

\NewDocumentCommand{\edv}{o}{\ensuremath{
\tilde{E}_{\dv[#1]
}}}

\NewDocumentCommand{\ndays}{o}{
\ensuremath{N_\mathrm{days}{\IfNoValueF{#1}{= {#1}}}}
}

\NewDocumentCommand{\se}{o}{\ensuremath{
S{\IfNoValueF{#1}{_{#1}}}
}}

\NewDocumentCommand{\ab}{o}{\ensuremath{
A{\IfNoValueF{#1}{_{#1}}}
}}

\NewDocumentCommand{\ahe}{s o O{=}}{\ensuremath{\ab[\he]
\IfNoValueF{#2}{
\IfNoValueTF{#3}{=}{#3}
\IfBooleanTF{#1}{#2}{#2 \%}}
}}

\NewDocumentCommand{\corr}{o}{\ensuremath{
\rho
\IfNoValueF{#1}{(#1)}
}}

\NewDocumentCommand{\xhel}{s o O{=} o}{\ensuremath{
\IfBooleanTF{#1}{\sigma_{c\IfNoValueF{#4}{,#4}}}{\abs{\sigma_{c\IfNoValueF{#4}{,#4}}}}
\IfNoValueF{#2}{
\IfNoValueTF{#3}{=}{#3}
#2}
}}

\NewDocumentCommand{\SpecInd}{o}{\ensuremath{\gamma
\IfNoValueF{#1}{_{#1}}}}
\NewDocumentCommand{\QT}{o}{\ensuremath{\mathrm{QT}
\IfNoValueF{#1}{= #1}}}

\NewDocumentCommand{\pten}{m m}{\ensuremath{
#1 \times 10^{#2}
}}

\NewDocumentCommand{\abs}{m}{\ensuremath{\left| #1 \right|}}

\NewDocumentCommand{\func}{m m o O{=}}{\ensuremath{#1\left(#2\right)\IfNoValueF{#3}{#4 #3}}
}

\newcommand{\rd}[1]{#1\textsuperscript{rd}}

\newcommand{\citepossessive}[1]{\citeauthor{#1}'s (\citeyear{#1})}

\newcommand{\degree}{\ensuremath{^\circ}}

\definecolor{C0}{HTML}{1f77b4}
\definecolor{C1}{HTML}{ff7f0e}
\definecolor{C2}{HTML}{2ca02c}
\definecolor{C3}{HTML}{d62728}
\definecolor{C4}{HTML}{9467bd}
\definecolor{C5}{HTML}{8c564b}

\definecolor{QTFitGreen}{HTML}{2ca02c}
\definecolor{DodgerBlue}{HTML}{1e90ff}
\definecolor{Fuchsia}{HTML}{ff00ff}
\definecolor{TabGreen}{HTML}{2ca02c}
\definecolor{Cyan}{HTML}{00ffff}
\definecolor{LimeGreen}{HTML}{32cd32}
\definecolor{Lime}{HTML}{00ff00}

\definecolor{MaxPink}{HTML}{e377c2}
\definecolor{MinPurple}{HTML}{9467bd}

\definecolor{q}{HTML}{228B22}
\definecolor{wc}{HTML}{FF8C00}
\definecolor{dnc}{HTML}{FF00FF}
\definecolor{todo}{HTML}{e13748}
\definecolor{ben}{HTML}{e13748}
\definecolor{bob}{HTML}{0080FF}

\NewDocumentCommand{\note}{s o m}{\IfBooleanF{#1}{\textcolor{q}{\textbf{Note}\IfNoValueF{#2}{ (#2)}: \textit{#3}}}}

\NewDocumentCommand{\question}{s o m}{\IfBooleanF{#1}{\textcolor{q}{\textbf{Q}\IfNoValueF{#2}{ (#2)}: \textit{#3}}}}
\NewDocumentCommand{\answer}{s o m}{\IfBooleanF{#1}{\textcolor{q}{\textbf{A}\IfNoValueF{#2}{ (#2)}: \textit{#3}}}}

\NewDocumentCommand{\wc}{s m}{\IfBooleanTF{#1}{#2}{\textcolor{wc}{\textbf{WC:} \textit{#2}}}}
\NewDocumentCommand{\ws}{s m}{\IfBooleanTF{#1}{#2}{\textcolor{wc}{\textbf{WS:} \textit{#2}}}}

\NewDocumentCommand{\delete}{s m}{\IfBooleanF{#1}{\textcolor{todo}{\textbf{Delete:} \textit{#2}}}}
\NewDocumentCommand{\todo}{s o m}{\IfBooleanF{#1}{\textcolor{todo}{\textbf{TODO}\IfNoValueF{#2}{ (#2)}: \textit{#3}}}}
\NewDocumentCommand{\verify}{s o m}{\IfBooleanTF{#1}{#3}{\textcolor{todo}{\textbf{VERIFY}\IfNoValueF{#2}{ (#2)}: \textit{#3}}}}
\NewDocumentCommand{\goal}{s o m}{\IfBooleanTF{#1}{#3}{\textcolor{todo}{\textbf{GOAL}\IfNoValueF{#2}{ (#2)}: \textit{#3}}}}

\NewDocumentCommand{\move}{s o m}{\textcolor{dnc}{\textbf{\IfBooleanTF{#1}{Duplicate}{Move}}\IfNoValueF{#2}{ (#2)}: \textit{#3}}}
\NewDocumentCommand{\dupe}{o m}{\move*[#1]{#2}}
\NewDocumentCommand{\intro}{s m}{\IfBooleanTF{#1}{\dupe[Intro]{#2}}{\move[Intro]{#2}}}
\NewDocumentCommand{\dnc}{s m}{\IfBooleanTF{#1}{\dupe[DnC]{#2}}{\move[DnC]{#2}}}
\NewDocumentCommand{\fw}{s m}{\IfBooleanTF{#1}{\dupe[Future Work]{#2}}{\move[DnC]{#2}}}

\DeclareDocumentCommand{\EmptyTimes}{O{black}}{\ensuremath{\mathord{\begin{tikzpicture}[line width=0.2ex, x=1.5ex, y=1.5ex]
\draw[color=#1] (0, 0.25) -- (0.25, 0.5) -- (0, 0.75) -- (0.25, 1.0) -- (0.5, 0.75) -- (0.75, 1.0) -- (1.0, 0.75) -- (0.75, 0.5) -- (1.0, 0.25) -- (0.75, 0) -- (0.5, 0.25) -- (0.25, 0) -- cycle;
\end{tikzpicture}}}}

\DeclareDocumentCommand{\SolidBand}{O{black} D{<}{>}{1}}{\ensuremath{\mathord{\begin{tikzpicture}[line width=1.25ex, x=1.25ex, y=1.25ex, yshift=5ex]
\draw[color=#1, opacity=#2] (0,0.5) -- (1.5,0.5);
\draw[opacity=0, line width=0.1ex] (0,0) -- (1.5,0);
\end{tikzpicture}}}}

\DeclareDocumentCommand{\SolidBandVertLines}{O{black} D{<}{>}{1}}{\ensuremath{\mathord{\begin{tikzpicture}[line width=1.25ex, x=1.25ex, y=1.25ex, yshift=5ex]
\draw[color=#1, opacity=#2] (0,0.5) -- (1.5,0.5);
\draw[opacity=0, line width=0.1ex] (0,0) -- (1.5,0);
\draw[color=#1, line width=0.2ex] (0, 0) -- (0, 1);
\draw[color=#1, line width=0.2ex] (1.5, 0) -- (1.5, 1);
\end{tikzpicture}}}}

\DeclareDocumentCommand{\SolidLine}{O{black}}{\ensuremath{\mathord{\begin{tikzpicture}[line width=0.3ex, x=1.25ex, y=1.25ex, yshift=5ex]
\draw[color=#1] (0,0.5) -- (1,0.5);
\draw[opacity=0] (0,0) -- (1,0);
\end{tikzpicture}}}}

\DeclareDocumentCommand{\DashedLine}{O{black} o D{<}{>}{1}}{\ensuremath{\mathord{\begin{tikzpicture}[line width=0.3ex, x=1.25ex, y=1.25ex, yshift=5ex]
\IfNoValueF{#2}{\draw[color=#2, opacity=#3] (0, 0.5) -- (1.75, 0.5);}
\draw[color=#1, opacity=#3] (0,0.5) -- (0.75,0.5);
\draw[color=#1, opacity=#3] (1.0,0.5) -- (1.75,0.5);
\draw[opacity=0] (0,0) -- (1.25,0);
\end{tikzpicture}}}}

\DeclareDocumentCommand{\HighlightDashedLine}{O{black} O{LimeGreen} D{<}{>}{1}}{\ensuremath{\mathord{\begin{tikzpicture}[line width=0.3ex, x=1.25ex, y=1.25ex, yshift=5ex]
\draw[color=#2, opacity=#3, line width=0.8ex] (0, 0.5) -- (1.75, 0.5);
\draw[color=#1] (0,0.5) -- (0.75,0.5);
\draw[color=#1] (1.0,0.5) -- (1.75,0.5);
\draw[opacity=0] (0,0) -- (1.25,0);
\end{tikzpicture}}}}

\DeclareDocumentCommand{\DotDotDotLine}{O{white} O{black}}{\ensuremath{\mathord{\begin{tikzpicture}[line width=0.3ex, x=1.25ex, y=1.25ex, yshift=5ex]
\draw[color=#1] (0, 0.5) -- (1.5, 0.5);
\draw[color=#2] (0,0.5) -- (0.3,0.5);
\draw [color=#2](0.6,0.5) -- (0.9,0.5);
\draw[color=#2] (1.2,0.5) -- (1.5,0.5);
\draw[opacity=0] (0,0) -- (1.5,0);
\end{tikzpicture}}}}

\DeclareDocumentCommand{\DotDotDotLineTwo}{O{yellow} O{black}}{\ensuremath{\mathord{\begin{tikzpicture}[line width=0.3ex, x=1.25ex, y=1.25ex, yshift=5ex]
\draw[color=#1] (0,0.5) -- (1.5,0.5);
\draw[color=#2] (0,0.5) -- (0.3,0.5);
\draw[color=#2] (0.6,0.5) -- (0.9,0.5);
\draw[color=#2] (1.2,0.5) -- (1.5,0.5);
\draw[opacity=0] (0,0) -- (1.5,0);
\end{tikzpicture}}}}

\DeclareDocumentCommand{\DashDotDotDotLine}{O{white} O{black}}{\ensuremath{\mathord{\begin{tikzpicture}[line width=0.3ex, x=1.25ex, y=1.25ex, yshift=5ex]
\draw[color=#1] (0, 0.5) -- (2.8, 0.5);
\draw[color=#2] (0,0.5) -- (1,0.5);
\draw[color=#2] (1.3,0.5) -- (1.6,0.5);
\draw[color=#2] (1.9,0.5) -- (2.2,0.5);
\draw[color=#2] (2.5,0.5) -- (2.8,0.5);
\draw[opacity=0] (0,0) -- (2.8,0);
\end{tikzpicture}}}}

\DeclareDocumentCommand{\Circle}{O{black}}{\ensuremath{\mathord{\begin{tikzpicture}[line width=0.3ex, x=1.25ex, y=1.25ex, yshift=5ex]
\draw[color=#1] circle (0.75ex);
\end{tikzpicture}}}}

\NewDocumentCommand{\sect}{o m}{Section~\ref{sec:#2}\IfNoValueF{#1}{ #1}}
\NewDocumentCommand{\eq}{o m}{\cref{eq:#2}\IfNoValueF{#1}{ #1}}
\NewDocumentCommand{\tbl}{o m}{\cref{tbl:#2}\IfNoValueF{#1}{ #1}}

  \usepackage{multirow}
\usepackage{tablefootnote}

\newcommand{\gridfigscale}{1}
\newcommand{\gridfigbasepath}{}

\NewDocumentCommand{\plotVswHist}{s}{
\IfBooleanTF{#1}{\begin{figure*}}{\begin{figure}}
\includegraphics[width=\linewidth]{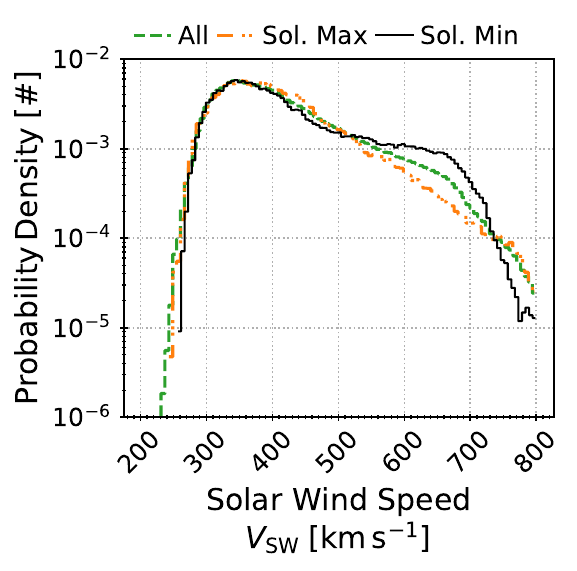}
\caption{\label{fig:vsw-hist}
Three probability density functions (PDFs) of the solar wind speed observed by the Wind Faraday cups at \au[1].
The PDFs indicate all the data observed (green dashed), data from solar maxima 23 and 24 (orange dash-dotted), and solar minima 23 and 24 (solid black).
}
\IfBooleanTF{#1}{\end{figure*}}{\end{figure}}
}

\NewDocumentCommand{\plotAheVsw}{s}{
\IfBooleanTF{#1}{\begin{figure*}}{\begin{figure}}
\includegraphics[width=\linewidth]{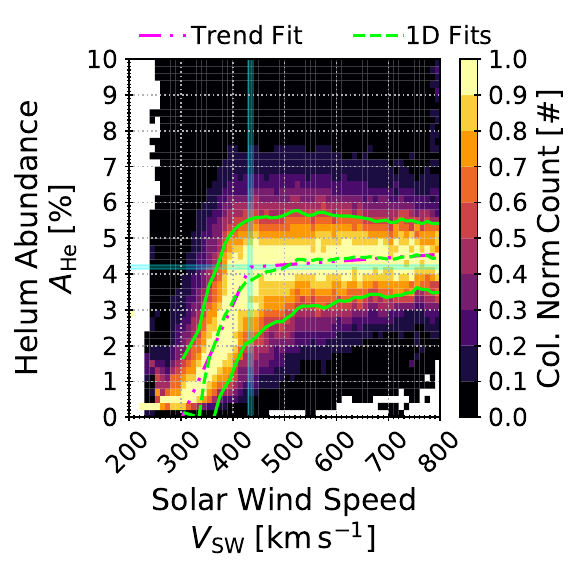}
\caption{\label{fig:ahe-vsw}
The helium abundance a function of solar wind speed.
\ahe\ has been normalized to its maximum value in each column.
The helium abundance monotonically increases from $0\%$ to $4.19\%$ in slow wind and saturates to this \ahe[4.19] in fast solar wind for which \vsw[433][>].
}
\IfBooleanTF{#1}{\end{figure*}}{\end{figure}}
}

\NewDocumentCommand{\plotXhelVsw}{s}{
\IfBooleanTF{#1}{\begin{figure*}}{\begin{figure}}
\includegraphics[width=\linewidth]{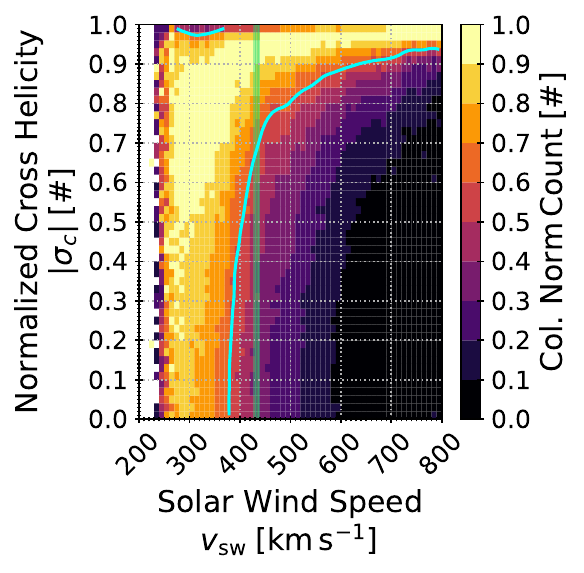}
\caption{\label{fig:xhel-vsw}
The cross helicity as a function of \vsw, normalized to its maximum value in each column.
The blue line indicates values at 60\% of the maximum in each column.
The green line indicates the saturation speed (\vs) as derived in \cref{fig:ahe-vsw}.
}
\IfBooleanTF{#1}{\end{figure*}}{\end{figure}}
}

\NewDocumentCommand{\plotVswSolMin}{s}{
\IfBooleanTF{#1}{\begin{figure*}}{\begin{figure}}
\includegraphics[width=\linewidth]{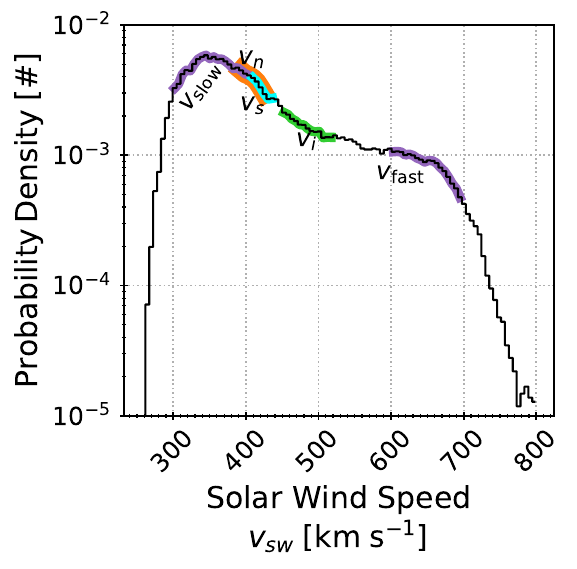}
\caption{\label{fig:vsw:SolMin}
The same PDF of \vsw\ in \cref{fig:vsw-hist}.
The highlighted purple, orange, blue, and green intervals are the same as in 
\cref{fig:ndens-vsw}: the fast and slow wind peaks derived from Gaussian fits (\vslow\ and \vfast\ in purple), the range of saturation speeds derived across \xhel\ (\vs\ in blue), the peaks of \n[\alpha] derived in \cref{fig:ndens-vsw} across \xhel\ (\vn\ in orange), and the speed at which the Gaussians used to derive \vslow\ and \vfast\ intersect (\vi\ in green).
}
\IfBooleanTF{#1}{\end{figure*}}{\end{figure}}
}

\NewDocumentCommand{\plotCartoon}{s}{
\IfBooleanTF{#1}{\begin{figure*}}{\begin{figure}}
\includegraphics[width=\linewidth]{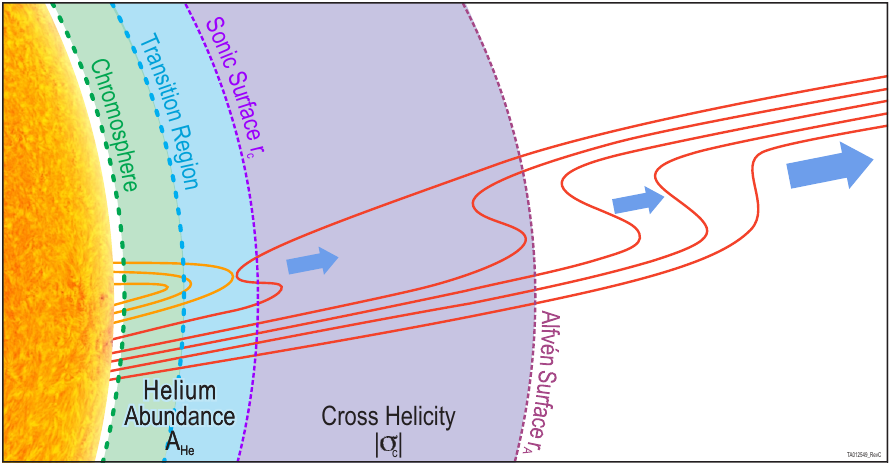}
\caption{\label{fig:cartoon}
A cartoon illustrating the relationship between the helium abundance (\ahe), cross helicity (\xhel), and magnetic field topology at the solar wind's source regions.
Closed magnetic loops are plotted in orange.
Open magnetic field lines are plotted in red.
The helium abundance is set below the sonic surface (\rc) in the chromosphere and transition region.
Between the sonic surface and the Alfvén surface (\rA), the cross helicity is set.
Above the Alfvén surface, the solar wind is magnetically disconnected from the Sun and \xhel\ can only decay.
The solar wind speed (blue arrows) increases during propagation through interplanetary space due to the decay of Alfvénic structures like switchbacks \citep{Bale2023,Raouafi2023,Rivera2024}.
Adapted from \citet{Akhavan-Tafti2024} Figure 3.
}
\IfBooleanTF{#1}{\end{figure*}}{\end{figure}}
}

\NewDocumentCommand{\plotSaturationFits}{s}{
\IfBooleanTF{#1}{\begin{figure*}}{\begin{figure}}
\includegraphics[width=\linewidth]{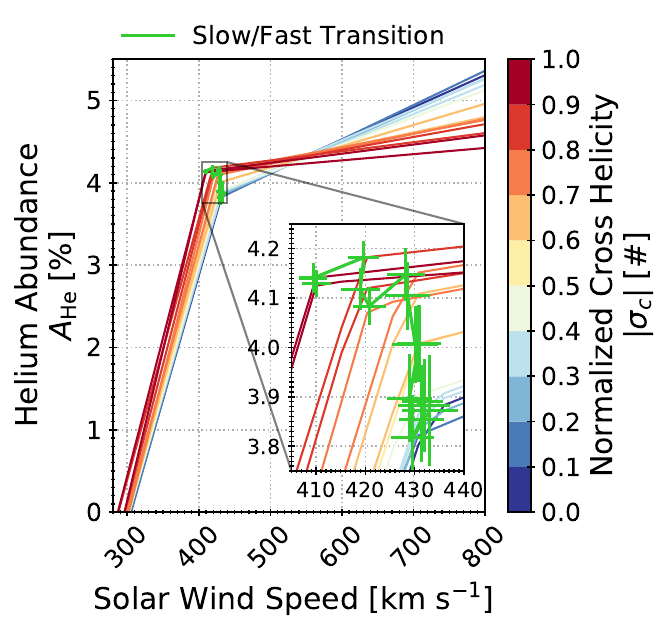}
\caption{\label{fig:sat-fits}
Fits to $\ahe\left(\vsw\right)$ in 15 \xhel\ quantiles, which are given by the color bar.
The green points are the fit values and uncertainties for the saturation points \satpoint.
The insert zooms in on the points.
}
\IfBooleanTF{#1}{\end{figure*}}{\end{figure}}
}

\NewDocumentCommand{\plotScaledSaturationFits}{s}{
\IfBooleanTF{#1}{\begin{figure*}}{\begin{figure}}
\includegraphics[width=\linewidth]{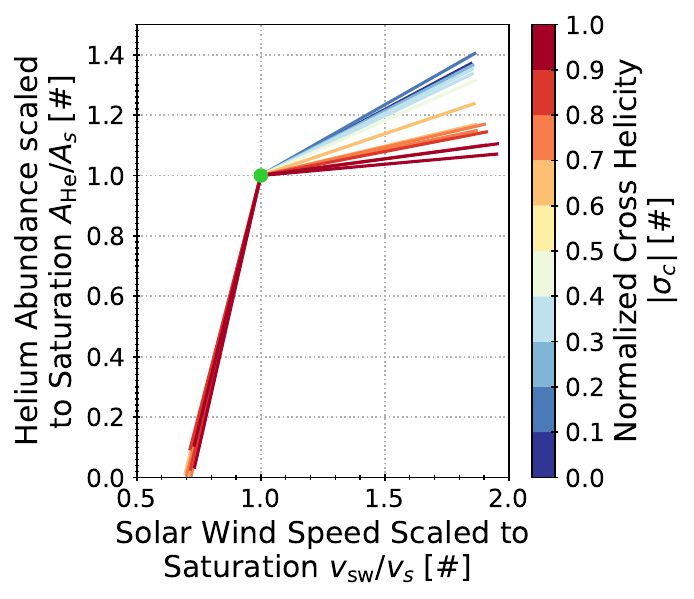}
\caption{\label{fig:sat-fits:scaled}
Fits to $\ahe\left(\vsw\right)$ in 15 \xhel\ quantiles, which are given by the color bar, rescaled to the saturation point \satpoint.
The green point indicates $(1, 1)$, the scaled saturation point.
}
\IfBooleanTF{#1}{\end{figure*}}{\end{figure}}
}

\NewDocumentCommand{\plotDualSatXhel}{s}{
\IfBooleanTF{#1}{\begin{figure*}}{\begin{figure}}
\includegraphics[width=\linewidth]{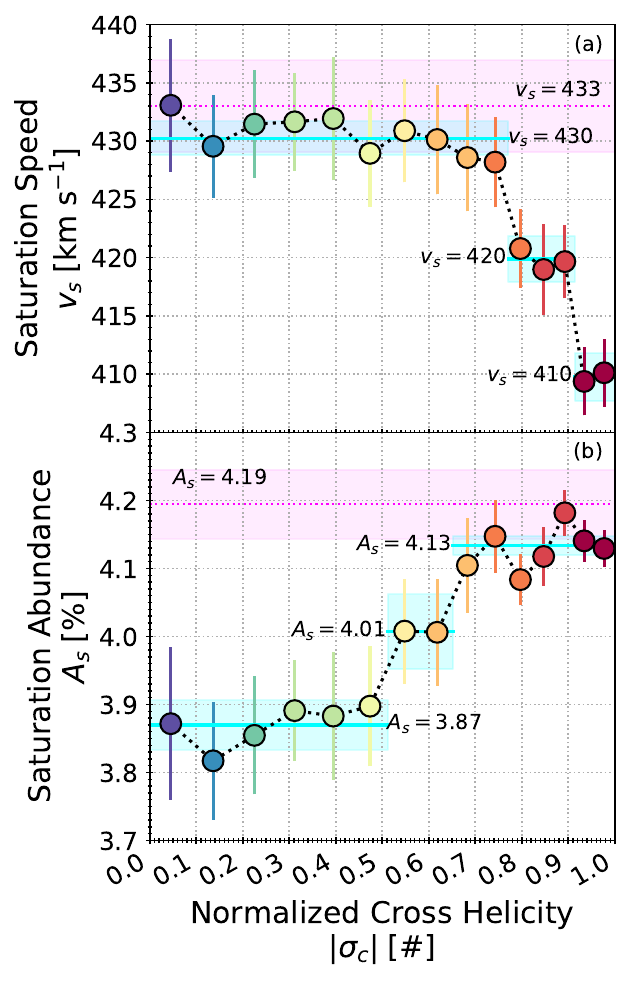}
\caption{\label{fig:dual_sat-xhel}
The \textbf{(a)} saturation speed (\vs) and \textbf{(b)} saturation abundance (\As) as a function of normalized cross helicity (\xhel).
Marker color indicates \xhel\ for visual comparison with \cref{fig:sat-fits,fig:sat-fits:scaled}.
The pink lines and shaded regions surrounding them are the values derived for all data in \cref{fig:ahe-vsw}.
The blue lines and shaded regions indicate the weighted mean and standard error of the mean for the indicated ranges of normalized cross helicity.
Markers and error bars are colored by \xhel\ for visual comparison with later figures.
}
\IfBooleanTF{#1}{\end{figure*}}{\end{figure}}
}

\NewDocumentCommand{\plotNdensVsw}{s}{
\IfBooleanTF{#1}{\begin{figure*}}{\begin{figure}}
\includegraphics[width=\linewidth]{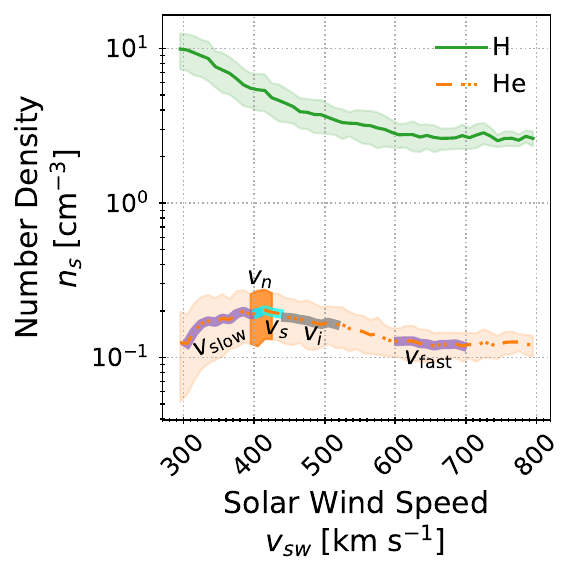}
\caption{\label{fig:ndens-vsw}
Mean alpha particle and proton number densities as a function of solar wind speed.
The semi-transparent regions are the standard deviations.
The highlighted regions on the \n[\He] trend indicate speeds within $1\sigma$ of the slow wind peak (\vslow) in \cref{fig:vsw-hist}, the fast wind peak (\vfast) in \cref{fig:vsw-hist}, and the saturation speed (\vs) derived in \cref{fig:ahe-vsw} for all data along with \vn, the peak of \n[\He] in this figure when the \n[\alpha] trend is recalculated across the \xhel\ quantiles.
}
\IfBooleanTF{#1}{\end{figure*}}{\end{figure}}
}

\NewDocumentCommand{\plotXhelAheVswContours}{s}{
\renewcommand{\gridfigscale}{1}
\renewcommand{\gridfigbasepath}{xhel-ahe-vsw}
\IfBooleanTF{#1}{\begin{figure*}}{\begin{figure}}
\begin{centering}
\includegraphics[
width=\gridfigscale\linewidth]{\gridfigbasepath/indicators-on-contourf}
\end{centering}
\caption{\label{fig:xhel-ahe-vsw:contours}
A contour plot of the solar wind speed (\vsw) as a function of normalized cross helicity (\xhel) and helium abundance (\ahe).
Contours at 355, 399, 407, 439, 450 and \kms[484] are drawn.
The first two are \vslow\ and the upper bound on it.
The middle two are the slowest and fastest \vs.
The latter two are the speed at which Gaussians fit to the slow and fast wind peaks during solar minima intersect and the lower bound on this intersection value.
All contours are smoothed with a $1\sigma$ filter for visual clarity.
}
\IfBooleanTF{#1}{\end{figure*}}{\end{figure}}
}

\NewDocumentCommand{\plotXhelAheVswContourArray}{s}{
\renewcommand{\gridfigscale}{\IfBooleanTF{#1}{0.475}{1}}
\renewcommand{\gridfigbasepath}{xhel-ahe-vsw/array-plots}
\IfBooleanTF{#1}{\begin{figure*}}{\begin{figure}}
\begin{centering}
\includegraphics[width=\gridfigscale\linewidth]{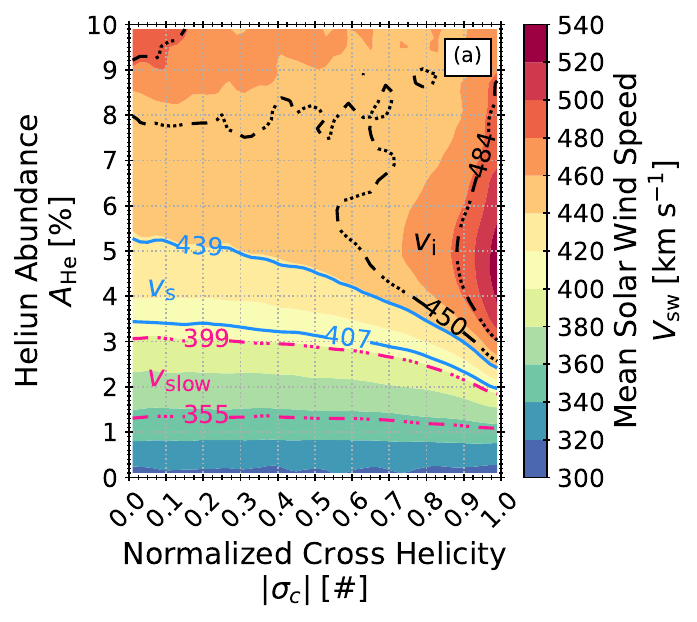}\\
\includegraphics[width=\gridfigscale\linewidth]{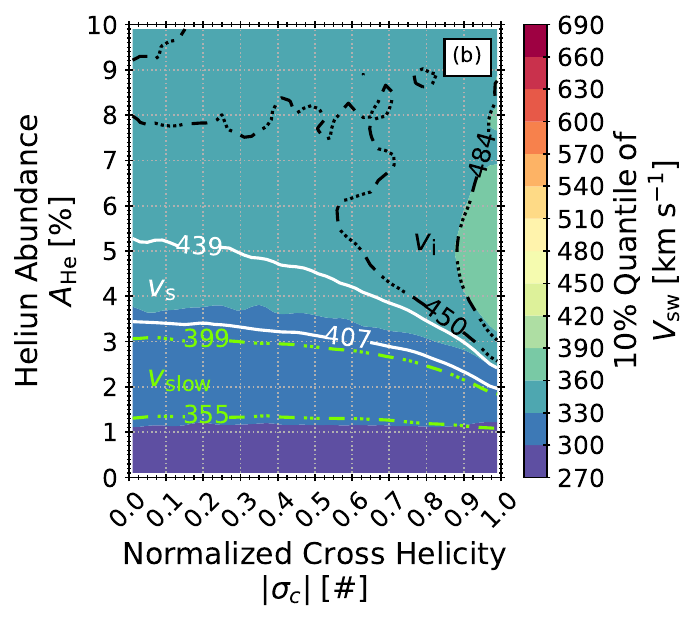}
\includegraphics[width=\gridfigscale\linewidth]{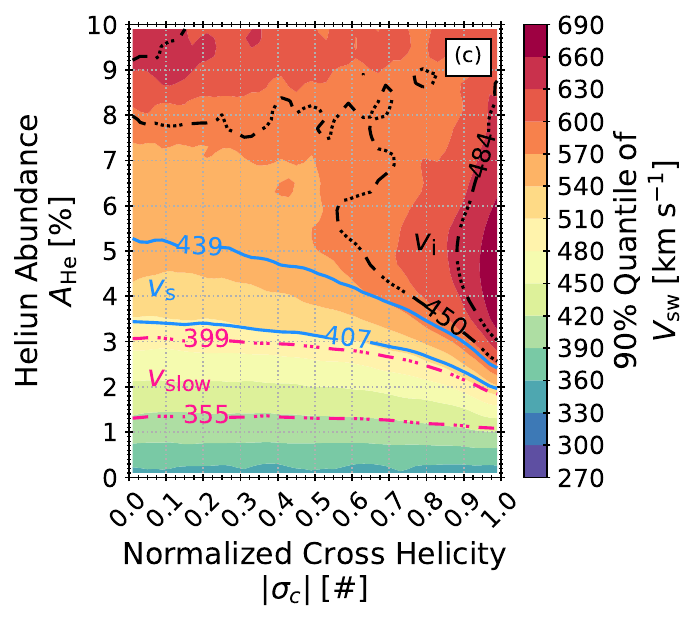}
\end{centering}
\caption{\label{fig:xhel-ahe-vsw:contours}
Contour plots of the solar wind speed (\vsw) as a function of normalized cross helicity (\xhel) and helium abundance (\ahe).
Panel (a) uses the mean \vsw.
Panels (b) and (c) use the 10\% and 90\% quantile of \vsw, respectively.
The color scale in Panels (b) and (c) is larger than the range in Panel (a).
Contours for mean \vsw*[355], 399, 407, 439, 450 and \kms[484] are drawn on all three panels.
The color of the contours depend on the panels and are chosen for contrast.
The first two are \vslow\ and the upper bound on it.
The middle two are the slowest and fastest \vs.
The latter two are the speed at which Gaussians fit to the slow and fast wind peaks during solar minima intersect and the lower bound on this intersection value.
The area between the \vslow, \vs, and \vi\ contour pairs are labeled with the corresponding $v_X$ text.
All contours are smoothed with a $1\sigma$ filter for visual clarity.
}
\IfBooleanTF{#1}{\end{figure*}}{\end{figure}}
}

\NewDocumentCommand{\plotXhelAheVswContoursCategorization}{s}{
\renewcommand{\gridfigscale}{1}
\renewcommand{\gridfigbasepath}{xhel-ahe-vsw/array-plots}
\IfBooleanTF{#1}{\begin{figure*}}{\begin{figure}}
\begin{centering}
\includegraphics[
width=\gridfigscale\linewidth]{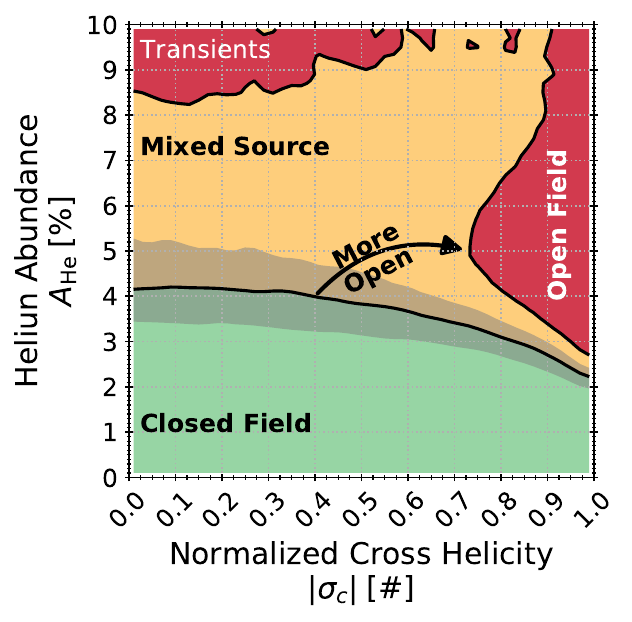}
\end{centering}
\caption{\label{fig:xhel-ahe-vsw:contours:categorization}
A suggestive cartoon derived from \cref{fig:xhel-ahe-vsw:contours} (a).
Contours indicating suggested regions of the \func{\relax}{\xhel,\ahe} plane.
Black contours are drawn at \vsw*[425] and \kms[460].
The bottom region indicates solar wind likely from closed field source regions.
The right side of the plane is solar wind likely from open field source regions.
The middle of the plane is solar wind that, when measured at \au[1], is from a mixture of sources.
A forthcoming publication demonstrates that high speed solar wind in the top left corner of the plane is from transients.
The gray region indicates the full range of saturation speeds \vs.
The arrow in the middle region suggests that the \emph{Mixed Source} region is more likely to contain solar wind from an open field region roughly along the direction of the arrow.
The region for \vs\ is grayed out to indicate that the exact \vsw\ at which this transition happens likely depends on the solar wind sample under consideration.
}
\IfBooleanTF{#1}{\end{figure*}}{\end{figure}}
}

\newcommand{\eqAhe}{
\begin{equation}\label{eq:ahe}
\ahe = 100 \times \frac{\n[\He]}{\n[\Hy]}.
\end{equation}
}

\NewDocumentCommand{\eqXhel}{s}{
\begin{equation}\label{eq:xhel}
\xhel* = \frac{e^+ - e^-}{e^+ + e^-}\IfBooleanF{#1}{.}
\end{equation}
}

\NewDocumentCommand{\eqdnn}{s}{
\begin{equation}\label{eq:dnn}
\abs{\frac{\delta n}{n}} = \abs{\frac{n - \langle n \rangle}{n}}.
\end{equation}
}

\newcommand{\twolinefcn}{
\begin{equation} \label{eq:two-line}
A(v) = \mathrm{min}\left[m_1(v - v_1), m_2 (v - v_2)\right]
\end{equation}
}

\newcommand{\eqVsat}{
\begin{equation}\label{eq:vs}
v_s = \frac{m_1 v_1 - m_2 v_2}{m_1 - m_2}.
\end{equation}
}

\clearpage{}

\newcommand{\DataTable}{
\begin{table}
\centering
\begin{tabular}{cccccc}
\hline\hline
 & $A_s$ &  $v_s$ &  $v_o$ &  $m_\mathrm{fast}$ \\
\xhel & $\left[\%\right]$ & $\left[\mathrm{km \; s^{-1}}\right]$ & $\left[\mathrm{km \; s^{-1}}\right]$ & $\left[\% \; \mathrm{km^{-1} \; s}\right]$ \\
\hline
0.05 &   $3.87 \pm 0.11$ &      $433 \pm 6$ &      $304 \pm 4$ &      $0.0039 \pm 0.0007$ \\
0.14 &   $3.82 \pm 0.09$ &      $430 \pm 4$ &      $305 \pm 3$ &      $0.0042 \pm 0.0004$ \\
0.22 &   $3.85 \pm 0.09$ &      $431 \pm 5$ &      $303 \pm 3$ &      $0.0038 \pm 0.0005$ \\
0.31 &   $3.89 \pm 0.07$ &      $432 \pm 4$ &      $302 \pm 4$ &      $0.0037 \pm 0.0004$ \\
0.39 &   $3.88 \pm 0.09$ &      $432 \pm 5$ &      $298 \pm 4$ &      $0.0035 \pm 0.0005$ \\
0.47 &   $3.90 \pm 0.09$ &      $429 \pm 5$ &      $303 \pm 4$ &      $0.0033 \pm 0.0004$ \\
0.55 &   $4.01 \pm 0.08$ &      $431 \pm 4$ &      $302 \pm 4$ &      $0.0026 \pm 0.0004$ \\
0.62 &   $4.01 \pm 0.08$ &      $430 \pm 5$ &      $297 \pm 5$ &      $0.0026 \pm 0.0004$ \\
0.68 &   $4.10 \pm 0.07$ &      $429 \pm 5$ &      $302 \pm 4$ &      $0.0019 \pm 0.0003$ \\
0.74 &   $4.15 \pm 0.05$ &      $428 \pm 4$ &      $299 \pm 4$ &      $0.0017 \pm 0.0002$ \\
0.80  &   $4.08 \pm 0.04$ &      $421 \pm 3$ &      $302 \pm 3$ &      $0.0018 \pm 0.0002$ \\
0.85 &   $4.12 \pm 0.04$ &      $419 \pm 4$ &      $297 \pm 4$ &      $0.0016 \pm 0.0002$ \\
0.89 &   $4.18 \pm 0.03$ &      $420 \pm 3$ &      $287 \pm 4$ &      $0.0011 \pm 0.0001$ \\
0.93 &   $4.14 \pm 0.03$ &      $409 \pm 3$ &      $296 \pm 3$ &      $0.0011 \pm 0.0001$ \\
0.98 &   $4.13 \pm 0.03$ &      $410 \pm 3$ &      $287 \pm 4$ &      $0.0008 \pm 0.0001$ \\
\hline
All  &   $4.19 \pm 0.05$ &      $433 \pm 4$ &      $302 \pm 4$ &      $0.0010 \pm 0.0002$ \\
Low  &   $3.87 \pm 0.04$ &      $430 \pm 1$ &         --- &       --- \\
Mid  &   $4.01 \pm 0.06$ &      $420 \pm 2$ &         --- &       --- \\
High &   $4.13 \pm 0.01$ &      $410 \pm 2$ &         --- &       --- \\
\hline\hline
\end{tabular}
\caption{\label{tbl:data}
Derived fit paramters as a function of normalized cross helicity \xhel.
The fit parameters are the saturation abundance (\As), the saturation speed (\vs), the speed $v_o$ at which the slow wind portion of the fit ($v < \vs$) intersects the x-axis, and the slope $m_\mathrm{fast}$ of the fast wind portion of the fit ($v > \vs$).
\emph{All} values are derived in \cref{fig:ahe-vsw} for all data.
\emph{High}, \emph{Mid}, and \emph{Low} are derived for the ranges indicated in \cref{fig:dual_sat-xhel} for \vs\ and \As.
As such, \emph{Low}, \emph{Mid}, and \emph{High} refer to different ranges in \xhel\ for these two quantities.
}
\end{table}
}\clearpage{}
\clearpage{}

\newcommand{\SpeedTable}{
\begin{table}
\centering
\begin{tabular}{cc}
\hline
\hline
Speed & Typical Value $\left[\mathrm{km \, s^{-1}}\right]$\\
\hline
\vfast & $622 \pm 58$ \\
\vi & $484 \pm 34$ \\
\vs & $407$ to $439$ \\
\vn & $409 \pm 15$ \\
\vslow & $355 \pm 44$ \\
\hline
\hline
\end{tabular}
\caption{\label{tbl:speeds}
Key speeds highlighted in \cref{fig:ndens-vsw}.
These are the speeds of the slow wind peak (\vslow) in \cref{fig:vsw-hist} for solar minima, the fast wind peak (\vfast) in \cref{fig:vsw-hist} for solar minima, the intersection of the Gaussians fit to these peaks (\vi), the saturation speed (\vs) derived in \cref{fig:ahe-vsw} for all data along with \vn, and the speed corresponding to the peak of \n[\He] in \cref{fig:ndens-vsw} when the \n[\He] trend is recalculated for each \xhel\ quantile.
\vs\ is the range of values (including uncertainties) in \cref{tbl:data}.
}
\end{table}
}\clearpage{}
 
\graphicspath{{./}{figures/}}
\DeclareGraphicsExtensions{.pdf,.png,.jpg,.eps,}

\received{\today}
\submitjournal{ApJ Letters}

\usepackage{lipsum}
\NewDocumentCommand{\BlindText}{O{6}}{\todo{Remove blind text. This is here to help figures render nicely.}
\textcolor{white}{\lipsum*[1-#1]}}

\NewDocumentCommand{\satpoint}{s o O{=}}{\ensuremath{\left(\vs,\As\right)
\IfNoValueF{#2}{
#3
\IfBooleanTF{#1}{#2}{left(#2\right)}}
}}

\begin{document}

\title{
Cross Helicity and the Helium Abundance as an \emph{in situ} Metric of Solar Wind Acceleration
%\\
%Characterizing the Transition from Magnetically Closed to Magnetically Open Solar Wind Sources and Identifying the Origin of the Alfvénic Slow Wind 
}

\shorttitle{Fast and Slow Wind: \vsw, \ahe, \& \xhel.}
\shortauthors{Alterman and D'Amicis}

\newcommand{\SwRI}{
\affiliation{Southwest Research Institute \\
6220 Culebra Road \\
San Antonio, TX 78238, USA}
}

\newcommand{\Goddard}{
\affiliation{Heliophysics Science Division \\
NASA Goddard Space Flight Center \\
8800 Greenbelt, RD\\
Greenbelt, MD 20771, USA}
}

\newcommand{\APL}{
\affiliation{Applied Physics Laboratory \\
The Johns Hopkins University \\
Laurel, MD 20723, USA}
}

\newcommand{\UTSA}{
\affiliation{Department of Physics and Astronomy\\
 University of Texas at San Antonio\\
 San Antonio, TX 78249, USA}
}

\newcommand{\IAPS}{
\affiliation{INAF - Institute for Space Astrophysics and Planetology\\
Via Fosso del Cavaliere, 100\\
00133 Rome, Italy}
}

\correspondingauthor{B.\ L.\ Alterman}
\email{b.l.alterman@nasa.gov}

\author[0000-0001-6673-3432]{B.\ L.\ Alterman}
\Goddard
\SwRI

\author[0000-0003-2647-117X]{Raffaella D'Amicis}
\IAPS

\begin{abstract}

The two-state solar wind paradigm is based on observations showing that slow and fast solar wind have distinct properties like helium abundances, kinetic signatures, elemental composition, and charge-state ratios.
Nominally, the fast wind originates from solar sources that are continuously magnetically open to the heliosphere like coronal holes 
while the slow wind is from solar sources that are only intermittently open to the heliosphere like helmet streamers and pseudostreamers.
The Alfvénic slow wind is an emerging \rd{3} class of solar wind that challenges the two-state fast/slow paradigm.
It has slow wind speeds but is highly Alfvénic, i.e.~has a high correlation between velocity and magnetic field fluctuations along with low compressibility typical of Alfvén waves, which is typically observed in fast wind.
Its other properties are also more similar to the fast than slow wind.
From 28 years of Wind observations at \au[1], we derive the solar wind helium abundance (\ahe), Alfvénicity (\xhel), and solar wind speed (\vsw).
Characterizing \vsw\ as a function of \xhel\ and \ahe, we show that the maximum solar wind speed for plasma accelerated in source regions that are intermittently open is faster than the minimum solar wind speed for plasma accelerated in continuously open regions.
We infer that the Alfvénic slow wind is likely solar wind originating from open-field regions with speeds below the maximum solar wind speed for plasma from intermittently open regions.
We then discuss possible implications for solar wind acceleration.
Finally, we utilize the combination of helium abundance and normalized cross helicity to present a novel solar wind categorization scheme that illustrates the transition in observations of solar wind at \au[1] from magnetically closed to magnetically open sources.

\end{abstract}


\section{Introduction \label{sec:intro}} 

The Sun's coronal plasma becomes the solar wind at the height where its speed transitions from sub- to supersonic as its thermal energy is converted to kinetic energy \citep{Parker1958a,Meyer-Vernet2007a}.
This height \rc\ is typically referred to as the ``sonic point'' and is considered to be at roughly \Rs[\sim5].
This is also where the solar wind reaches the minimum speed required to escape the Sun's gravitational pull.
However, this energy conversion mechanism does not provide sufficient energy for the solar wind to reach the asymptotic, fastest non-transient speeds observed at \au[1].
Additional energy must be deposited into the solar wind for it to reach these speeds  \citep{Leer1980,Hansteen2012,Holzer1981,Holzer1980a,Johnstone2015}.
Alfvén waves, including switchbacks, are thought to be one key source of such energy \citep{PSP:SWEAP:1,PSP:FIELDS:1,Balogh1999c,Larosa2021,Jagarlamudi2023,Huang2023,Rivera2024}.

\plotVswHist
Broadly, there are two types of solar wind sources at the Sun: magnetically open and magnetically closed \citep{Poletto2013,Gosling1997}.
Sources like coronal holes (CH) are magnetically open to the heliosphere such that the magnetic fields are radial \citep{Phillips1994,Geiss1995}.
Sources like helmet streamers, pseudostreamers, and the boundaries between pseudostreamers and CHs have magnetic field topologies that are more complex and, though intermittently open to the heliosphere, are often referred to as magnetically closed \citep{Fisk1999,Subramanian2010,Antiochos2011,Crooker2012,Abbo2016,Antonucci2005}.
We will primarily use ``closed'' terminology.
During solar minima, magnetically open sources are confined to the Sun's polar regions and solar wind sources in the Sun's equatorial regions are considered closed.

The latitudinal stratification of magnetically open and closed sources during solar minima leads to a bimodal solar wind distribution observed at \au[1] during these time periods \citep{Bavassano1991}.
The bimodal structure is mostly absent during solar maxima and is largely obscured when all the data is plotted \citep{Bavassano1991}.
\cref{fig:vsw-hist} illustrates this.
It contains three probability distributions of the solar wind speed (\vsw) observed by the Wind Faraday cups when the spacecraft was outside of Earth's magnetosphere, one each for all the data observed, data from solar minima 24 and 25, and data from solar maxima 23 and 24.
We define these intervals following \citet{DAmicis2021}.
The bimodal nature of \vsw's distribution during solar minima has motivated a two-state fast/slow solar wind paradigm with the transition occurring  somewhere between \kms[400] and \kms[600] \citep{Schwenn2006,Fu2018}, though the exact value is typically chosen in an \emph{ad hoc} fashion.
\citet{Wind:SWE:Wk} report that the peaks of the slow and fast wind probability density functions (PDFs) collected during solar minima are \vslow[355\pm44] and \vfast[622\pm58].
Sources with magnetically open topologies accelerate fast wind and sources with magnetically closed topologies accelerate slow wind \citep{Baker2023,Arge2003,Arge2013,Wang1990,Arge2000}, which is confined to the heliospheric current sheet during solar minima \citep{Schwenn2006}.

Under the two-state solar wind paradigm, there are many differences observed between the fast and slow solar wind.
For example, the fast and slow wind have different kinetic properties \citep{Kasper2008,Kasper2017,Tracy2016,Wind:SWE:bimax,Stakhiv2016,Alterman2018,Berger2011,Klein2021,Verniero2020,Verniero2022,Durovcova2019}.
They also display different chemical abundances and charge-state ratios \citep{vonSteiger2000,Geiss1995,Geiss1995b,Zhao:InSituComposition:Sources,Zhao2022,Xu2014,Fu2017,Fu2015,Ervin2023,Brooks2015}.
However, classifying the solar source region of a given \textit{in situ} solar wind observation by its speed along is known to be inaccurate.
For example, observations indicate that there is a subset of solar wind with speeds characteristic of slow wind but other properties regularly associated with fast wind from magnetically open sources \citep{DAmicis2021a,DAmicis2021,DAmicis2018,Damicis2016,DAmicis2015,Yardley2024}.
Such wind was first identified in a case study by \citet{Marsch1981} and then on a statistical basis by \citet{DAmicis2011a}.
It is typically referred to as the ``Alfvénic slow wind''.

Observational case studies have identified individual sources of ASW that are mutually consistent to various degrees and supported by various modeling and theoretical works, but a statistical determination of the source of the ASW is yet to be proposed.
Key theories and observations suggest that the Alfvénic slow wind is from the low speed extension of sources typically associated with fast wind \citep[e.g.][]{DAmicis2015}.
For example, Alfvénic slow wind may be related to small coronal holes at low solar latitudes that replace polar coronal holes during solar maximum and are predominant during this phase of the solar cycle \citep{Wang1994a,Wang2019}.
\cite{Panasenco2019,Panasenco2020} have also identified source regions with strong non-monotonic expansion in the low corona below \Rs[1.6].
Such regions are usually found in the neighborhood of large-scale pseudostreamers.
Indeed, the topology of pseudostreamers allows for the formation and development of twin filament channels, a magnetic configuration that creates conditions for a strong divergence of the pseudostreamer open magnetic field \citep{Panasenco2013,Panasenco2019}. 
The strong divergence of the magnetic field as well as the non-monotonic expansion in the low corona would decelerate fast solar wind, setting the conditions for the development of the Alfvénic slow wind \citep{Panasenco2020}.
Recent studies have shown some evidence of Alfvénic slow wind intervals originating from narrow open-field corridors close to an active region related to separatrices or quasi-separatrix layers as in the S-web model \citep{Baker2023} or more in general originating from open-closed magnetic field boundaries such as active region edges \citep{Yardley2024,Ervin2024} or the quiet Sun \citep{Wang2019c} released into the heliosphere by interchange reconnection.

The helium abundance in the Alfvénic slow wind is more similar to the typical fast wind than the traditional slow wind.
Solar wind ions are composed of $\sim$95\% protons or ionized hydrogen (\Hy) and $\gtrsim4\%$ alpha particles or fully ionized helium (\He), with the remainder made of heavier elements.
Solar wind helium was first observed by \emph{Mariner 2} \citep{Mariner2:Ahe:1,Mariner2:Ahe:2}.
The solar wind helium abundance (\ahe) is given by the alpha-to-proton density ratio in units of percent
\eqAhe
The helium abundance monotonically increases from vanishingly small values to \ahe[4][\approx] in the slow wind from \kms[\sim250] to \kms[\sim400] and remains approximately constant at \ahe[4][\approx] for faster speeds \citep{Kasper:Ahe,Alterman2019,Aellig:Ahe,Ogilvie1974,Yogesh:Ahe,Fu2018}.
The change of \ahe's gradient as a function of \vsw\ at \kms[\sim \! 400] is one justification for the two-state paradigm, in which slow wind has speeds \vsw[400][\lesssim] and fast wind has speeds \vsw[400][\gtrsim].
\cref{sec:analysis:ahe,fig:ahe-vsw} quantify this statement.
The helium abundance is likely driven by the coronal heat flux into the chromosphere/transition region and local changes in the magnetic topology change how the coronal heat flux propagates into these regions \citep{Lie-Svendsen2001,Lie-Svendsen2002,Endeve2005,Lie-Svendsen2003,Hansteen1997}.
Because the sonic critical point is above the chromosphere and transition region, \ahe\ likely reflects source region properties below \rc.

\plotCartoon*
The first solar wind observations by \emph{Mariner 5} showed that fluctuations in the solar wind's magnetic field (\BB) components and velocity (\Bv) components are highly correlated while the density fluctuations were minimal ($\delta n/n \approx 0$) \citep{Belcher1969,Belcher1971}.
Under magnetohydrodynamic (MHD) theory, pure such correlations are indicate of Alfvén waves \citep{Alfven1942,Alfven1943}.
This condition is also referred as ``weak compressibility''  and involves fluctuations in the magnitude of the total magnetic field that are found to be much smaller than the magnetic field fluctuations  \citep{Bruno2001,Matteini2015}.
Later solar wind observations collected by the \emph{Helios} spacecraft in the inner heliosphere showed that Alfvénic correlations are markedly stronger within the main portion of fast streams, while they are weak in intervals of slow wind \citep{Tu1995,LR:turbulence}.

The Alfvénicity, or level of such \BB-\Bv\ correlations, is typically measured by the normalized cross helicity \citep{Tu1995,LR:turbulence,Woodham2018}.
The normalized cross helicity is given by
\eqXhel
Here, $e^\pm = \frac{1}{2}\langle \left(z^\pm\right)^2\rangle$ are the energies associated with the Elsässer variables, which are $\Bz^\pm = \Bv \pm \frac{1}{\sqrt{\mu_0\rho}}\Bb$ for velocity \Bv, magnetic field \Bb, and solar wind mass density $\rho$ \citep{ElsasserVariables,Tu1989,Grappin1991}.
The Elsässer variables are defined in such a way that $\Bz^+$ always corresponds to modes propagating away from the Sun while $\Bz^{-}$ corresponds to modes towards it.
The closer \xhel* is to $\pm1$, the more the predominant mode reflects pure Alfvénic fluctuations.
As shown in \cref{fig:xhel-vsw}, slow wind observations do not display a preferred \xhel, while fast wind \xhel\ tends towards 1 for \vsw*[\vs][\gtrsim].
Helios observations show that, although \xhel\ decays with distance from the sun \citep{Tu1992}, the decay most strongly impacts low cross helicity wind \citep{Marsch1990}.

Observations from Parker Solar Probe (Probe) during solar minimum 25 show that the near-Sun solar wind was highly Alfvénic, irrespective of its speed \citep{Raouafi2023}.
Highly Alfvénic solar wind is often full of switchbacks \citep{PSP:SWEAP:1,PSP:FIELDS:1,Bourouaine2020,McManus2020}.
Switchbacks are large scale rotations in the solar wind's magnetic field in the absence of changes in the magnetic field magnitude and they originate with interchange reconnection.
Broadly, interchange reconnection releases coronal plasma into the solar wind \citep{Fisk2001,Fisk2005,Fisk2020,Wyper2022,Bale2023}, producing magnetic structures called jets \citep{Raouafi2023}.
These magnetic structures evolve between the sonic and Alfvén surfaces \citep{Bale2021,Bale2023,Touresse2024,Drake2021,Zank2020}, becoming switchbacks just above or near the Alfvén surface \citep{Akhavan-Tafti2024}.
The deposition of energy from such structures during the solar wind's propagation through interplanetary space \citep{DudokDeWit2020,Tenerani2021,Rasca2021} accelerates the solar wind to its fastest, non-transient speeds observed below the orbit of Venus and near-Earth \citep{Rivera2024,Soni2024,Wind:SWE:Wk}.

The Alfvén radius (\rA) is the distance from the Sun above the solar wind is magnetically disconnected from the Sun and motion on the Sun's surface (e.g. from interchange reconnection) can no longer modify the solar wind.
The Alfvén radius is above the sonic point at a nominal height of \rA[20][\sim] \citep{PSP:SWEAP:Ra}.
\citet{Akhavan-Tafti2024} show that switchbacks are not observed below the Alfvén radius.
This means that the solar wind's Alfvénicity is likely set at heights near the Alfvén radius, which is above the sonic critical point and above the heights where \ahe\ is set.
How switchbacks evolve during propagation through interplanetary space likely depends on the magnetic topology of the solar wind's source region and lead to the differences in \xhel\ observed near-Earth.
At \au[1], these differences manifest in differences in \xhel\ as a function of \vsw.
As with \ahe, \xhel\ is typically $> 0.6$ for \vsw[400][\gtrsim], while slow wind does not display a preferred \xhel\ \citep{Tu1992a,DAmicis2021}.
In other words, a larger \xhel\ implies that it is more likely a given parcel of solar wind originated from a solar source region with a magnetically open topology.

\cref{fig:cartoon} is a cartoon illustrating this relationship between \xhel, \ahe, and magnetic field topology in the solar wind's source regions on the sun.
The orange and red lines indicate closed and open magnetic field lines, respectively.
The helium abundance is set below the sonic surface (\rc) in the chromosphere and transition region.
The cross helicity is set between the sonic surface and the Alfvén surface (\rA).
Above the Alfvén surface at heights $r > \rA$, the solar wind is magnetically disconnected from the Sun and \xhel\ can only decay.
Alfvén waves in the solar wind can manifest as switchbacks.
The decay of Alfvénic structures like switchbacks accelerates the solar wind to the fastest, non-transient speeds observed near \au[1] \citep{Bale2023,Raouafi2023,Rivera2024,Wind:SWE:Wk,Soni2024}.
The increase in \vsw\ during propagation through interplanetary space is indicated by the increase in size of the blue arrows with increasing distance from the Sun.
As this figure illustrates, the combination of \xhel\ and \ahe\ help identify solar wind source regions based on physical processes that occur above and below the sonic critical point.

We utilize 28 years of Wind Faraday cup \citep{Wind:SWE} and magnetic field observations \citep{Wind:MFI:Lepping,Wind:MFI} to investigate the relationship between \ahe, \xhel, and \vsw.
The time period covers 1994 to 2022.
\cref{fig:dual_sat-xhel} shows that the speed (\vs) observed near \au[1] at which the dominant source of the solar wind in the changes from magnetically closed to magnetically open decreases as the Alfvénicity increases while the \ahe\ characteristic of this transition increases with Alfvénicity.
This more nuanced characterization of the solar wind's origin helps explain the source of the Alfvénic slow wind, i.e. why some slow solar wind behaves more similarly to what is normally characterized as fast wind.
From these observations, we then infer that \He\ in open field source regions is accelerated in a manner similar to \Hy, while there is insufficient energy to continuously accelerate \He\ into the solar wind in closed field regions.
In such closed regions, \He\ either serves as a free energy source that is drained so that \Hy\ can be accelerated in closed field regions or it is not energized in the first place.
We then show that this careful determination of solar sources leads to a natural categorization of solar wind speed as a function of \ahe\ and \xhel.
Such a classification scheme is significant because it maps \textit{in situ} solar wind observations to their source regions and the associated magnetic topologies without elemental and charge state composition observations, which require mass spectrometers to collect.
Combining this classifications scheme with the different speeds derived in this paper, we characterize different regions of the bimodal distribution of solar wind speeds observed during solar minima near Earth.

\section{Observations \label{sec:obs}} 

Solar wind measurements are derived from Wind Faraday cup observations by fitting the measured solar wind charge flux with a model velocity distribution function (VDF).
Multiple datasets, each using a different model VDF have been produced \citep{Wind:SWE:bimax,Wind:SWE:dvapbimax,Alterman2018}.
We use the Wind/FC observations derived by fitting them to a model distribution function that accounts for a bi-Maxwellian \He\ population and a single population of bi-Maxwellian \Hy\ \citep{Wind:SWE:bimax}.
We select our data based on the following requirements.
\begin{compactenum}
\item Wind is outside of Earth's magnetosphere.\footnote{See Figure 1 of \citet{Wilson2021}. The black circle is the Moon’s orbit. Even in the first years of the mission, the spacecraft went beyond the Moon’s orbit and was therefore necessarily outside of the magnetosphere.}
\item The magnetic field co-latitude is within 65\degree of the ecliptic.
\item The magnetic field fluctuations were not extreme over the $\sim$92s during which a given FC spectrum is collected.
\item The fitting routines return physically meaningful solar wind speeds along with both \Hy\ and \He\ densities.
\end{compactenum}
The data are hosted by NASA on CDAWeb.
The Wind magnetic field observations are provided by the Magnetic Field Investigation (MFI).
We use the version downsampled to the Wind/FC measurement time and provided in the Wind/SWE data files.

\section{Analysis\label{sec:analysis}}

\subsection{Helium Abundance \label{sec:analysis:ahe}}
\plotAheVsw
\cref{fig:ahe-vsw} is a 2D histogram of the helium abundance \ahe\ as a function of \vsw\ over the range 200 to \kms[800].
Because Wind observes slow solar wind more often than fast wind, we have normalized the occurrence rate in each column to its maximum value so that the trend of \ahe\ with \vsw\ is not obscured by the sampling frequency apparent in \cref{fig:vsw-hist}.
By inspection, it is clear that \ahe\ increases from $\sim0$ to $\gtrsim 4\%$ from the slowest observed solar wind up to \kms[\sim400].
Above these speeds, \ahe\ saturates at 4\% to 5\%.
Such behavior has been previously reported \citep[e.g.][]{Kasper:Ahe,ACE:SWICS:FSTransition}.

Next, we aim to quantify trend of \ahe\ as a function of \vsw\ and to determine the speed at which \ahe\ indicates a transition from slow to fast wind.
As we are concerned with the central behavior of \ahe\ as a function of \vsw, we reduce the impact of the large, asymmetric \ahe\ tails by fitting \ahe\ in each \vsw\ column with a Gaussian distribution and limiting the fits to values within 90\% of each column's maximum.
Here, we only consider speeds in the range 300 to \kms[800] because \cref{fig:ahe-vsw} indicates that \ahe\ is vanishingly small at lower speeds and likely correspond to the very slow solar wind, which may uniquely originate from the heliospheric plasma sheet \citep[HPS,][]{Sanchez-Diaz2016a}.
In \cref{fig:ahe-vsw}, the mean of the 1D distributions is plotted as a green dashed line; their $1\sigma$ uncertainties are given as solid green lines.
As would be expected from normally distributed values, these solid green lines are roughly within the 60\% to 70\% of maximum level in each column.
A similar technique to reduce the impact of distribution tails has been used to characterize the decay of alpha particle and proton beam differential flow with increasing Coulomb collision \citep{Alterman2018}. 

We then fit the mean and standard deviation from these 1D Gaussians with the minimum of two lines
\twolinefcn
using the standard deviations as the weights.
$A(v)$ is the helium abundance; $m_i$ is a given line's slope; and $v_i$ is its x-intercept.
\citet{Kasper:Ahe} refer to $v_i$ of the slower speed interval with the steeper gradient as the vanishing speed (\vv).
The intersection of these two lines gives the speed at which \ahe\ saturates to its fast wind value
\eqVsat
We parameterize this function to determine \vs, \As, \vv, and $m_\mathrm{fast}$.
These are the speed (\vs) and abundance (\As) where the gradient of \ahe\ changes along with the x-intercept of the line at speeds $v < \vs$ (\vv) and the slope at speeds $v > \vs$ ($m_\mathrm{fast}$).
We refer to \vs\ and \As\ as the saturation speed and saturation abundance.
Fitting the minimum of two lines in such a fashion has been used to identify quiet times in suprathermal observations by ACE/ULEIS \citep{STQT:selection-abundances}.
In \cref{fig:ahe-vsw}, the fit is plotted as a pink dash-dotted line that is roughly co-located with the green dashed line indicating the mean values from the 1D fits.
\cref{tbl:data} summarizes the fit parameters for this and later fits.
The blue vertical line is the saturation speed \vs[433 \pm 4.0], above which \ahe[4.19 \pm 0.05].

\citet{Wind:SWE:Wk} have fit the fast and slow wind distributions of \vsw\ observed during solar minima each with a Gaussian and determined their intersections.
Because they represent the extreme cases of ranges typically chosen to differentiate between fast and slow solar wind, they consider slow wind to have \vsw[400][\leq] and fast wind to have \vsw[600][\geq].
These speeds are \vslow[355 \pm 44] and \vfast[622 \pm 58].
These Gaussians intersect at \vi[484 \pm 34], which is 13 to \kms[89] faster than \vs\ in \cref{fig:ahe-vsw}.
\cref{tbl:speeds} summarizes \vslow, \vfast, \vi, and other significant speeds derived in this work.

\subsection{Normalized Cross Helicity \label{sec:analysis:xhel}}
\plotXhelVsw
The normalized cross helicity is a tool to measure the solar wind's Alfvénicity, i.e. how correlated fluctuations in the solar wind velocity and magnetic field are.
It is given by \cref{eq:xhel}.
The sign of \xhel* indicates propagation towards or away from the Sun.
Observations show that the fluctuations are predominantly outward from the Sun \citep{LR:turbulence}.
We have calculated it on a 1-hr time scale, a typical scale of Alfvénic fluctuations \citep[e.g.][]{Marsch1990,DAmicis2015}, and only consider the absolute value because we are concerned with the Alfvénic content of the solar wind and not the directionality of the fluctuations.

\cref{fig:xhel-vsw} is a 2D histogram of \xhel\ as a function of \vsw\ normalized as in \cref{fig:ahe-vsw} to account for the same observation frequency concerns as \cref{fig:ahe-vsw}.
\citet{DAmicis2021} have produced similar plots for different phases of solar activity that include the sign of \xhel\ that show distinct fast and slow wind behavior.
Their distribution of data during solar minima is dominated by low \xhel* in slow wind and their fast wind peak is predominantly \xhel\ approaching 1.
During solar maxima, they show two distinct slow wind peaks at low and high \xhel*.
The distribution of data in \cref{fig:xhel-vsw} superimposes these various peaks because it mixes all phases of solar activity.
The range of observed \xhel\ decreases as \vsw\ increases, as highlighted by the blue line, which indicates bins at 60\% of the maximum in each column.
This contour intersects the x-axis at \vsw[380].
In the slow wind with \vsw[380][\lesssim], the observations cover all possible \xhel.
The vertical green line indicates \vs\ including the $\pm \kms[4]$ uncertainty derived from \cref{fig:ahe-vsw}.
This line intersects the 60\% contour at $\xhel\ = 0.7$.
In fast wind with $\vsw > \vs$, bins within 60\% of the maximum in each column are limited to $\xhel\ > 0.7$.

\subsection{Combining \ahe\ and \xhel \label{sec:analysis:ahe-xhel}}
\plotSaturationFits
To characterize the relationship between \ahe, \xhel, and \vsw, we repeated the analysis in \cref{fig:ahe-vsw} for 15 quantiles in \xhel.
We use quantiles instead of uniform length intervals so that there is an equal number of observations in each \xhel\ interval.
As in \cref{fig:ahe-vsw}, we only consider speeds \vsw[300][\geq].
We also require that the $1\sigma$ fit uncertainty for \ahe\ in each \vsw\ column be less than 3 percentage points, thereby excluding \vsw\ bins for which the data is too sparse.
This primarily excludes bins in the fastest wind where the data is limited due to the sample frequency of the fastest solar wind.
\cref{tbl:data} provides the fit parameters, their uncertainties, and averages over intervals defined below.

\cref{fig:sat-fits} plots the resulting fits.
The line colors indicated \xhel\, which is given by the color bar.
The saturation points \satpoint\ are plotted in green and highlighted in the insert axis.
This insert axis has the same aspect ratio as the larger, parent axis.
Two broad groups of observations stand out, one about the gradients at speeds above and below \vs\ and the other about how the saturation point \satpoint\ itself varies with \xhel.

\plotScaledSaturationFits
First, the gradients of \As\ as a function of \vsw\ are larger at speeds below the saturation point \vs\ than above it.
These gradients are similar across all \xhel\ for speeds \vsw*[\vs][<] and the speed at which this component of the fit reaches \ahe*[0] is approximately constant at \kms[302 \pm 4].
For speeds \vsw*[\vs][>], the gradients decrease as \xhel\ increases. In other words, the change in gradients across \vs\ increases with \xhel.
Quantitatively, \grad[\ahe][\vsw] for \vsw*[\vs][>] monotonically decrease from approximately $\pten{4.2}{-3} \, \% \, \mathrm{km^{-1} \, s}$ to $\pten{0.8}{-3} \, \% \, \mathrm{km^{-1} \, s}$, and 82\% change.
\cref{fig:sat-fits:scaled} highlights these observations of the gradients by scaling the fits to their saturation points \satpoint, which is indicated by a green dot.
It shows that \grad[\ahe][\vsw] for \vsw*[\vs][>] has two groups such that for \xhel[0.68][\geq], \grad[\ahe][\vsw] drops by a factor of $\sim 0.27$ from its low \xhel\ value.

Second, as highlighted by the insert axis, \As\ and \vs\ are anti-correlated: \As\ increases with increasing \xhel\, while \vs\ decreases with increasing \xhel.
Moreover, the typical \ahe\ as a function of \vsw\ in non-Alfvénic wind with small \xhel\ is larger than the typical \ahe\ in Alfvénic solar wind at speeds \vsw[525][\gtrsim].
This difference in \ahe\ between low and high cross helicity solar wind increases with increasing \vsw[525][\gtrsim].

\plotDualSatXhel
\cref{fig:dual_sat-xhel} analyzes the saturation values \vs\ and \As\ as a function of \xhel.
Marker color indicates \xhel\ and matches \cref{fig:sat-fits,fig:sat-fits:scaled}.
The pink dashed lines and shaded regions surrounding them are the saturation values derived using all the data in \cref{fig:ahe-vsw}.
Although the relevant ranges of \xhel\ for \vs\ and \As\ are not equivalent, \vs\ and \As\ show distinct groupings across \xhel, which we indicate in blue.
The lines denote the weighted mean across the interval and shaded regions are the standard errors of the mean.

\cref{fig:dual_sat-xhel} (a) plots the saturation speed \vs\ as a function of \xhel.
It shows that, over all \xhel\ quantiles, \vs\ drops from $433 \pm 6$ to \kms[410 \pm 3], a \kms[23] or $5\%$ change.
However, there is a clear change in behavior at $\xhel\ = 0.7$.
From $\xhel\ = 0$ to $\xhel\ < 0.77$, \vs\ drops by \kms[5] from 433 to \kms[428].
This entire change is within the $1\sigma$ fit uncertainties for \vs\ in each of these $\xhel\ < 0.77$ quantiles.
As such, we consider \vs\ constant in this range of $\xhel\ < 0.77$ and take it to be \vs[430 \pm 1], the weighted average of the derived saturation speeds over this range.
For cross helicities $\xhel\ \geq 0.77$, \vs\ decreases from \kms[421] to \kms[410], a \kms[11] change.
Within this range, there are two groups of \xhel\ that are within their mutual uncertainties 
For $0.77 < \xhel\ \leq 0.91$ and $0.91 < \xhel$, we consider \vs[420 \pm 2] and \vs[410 \pm 2], respectively.

\DataTable
\cref{fig:dual_sat-xhel} (b) plots the saturation abundance \As, i.e. the abundances corresponding to \vs, as a function of \xhel.
\As\ increases with \xhel\ and low cross helicity \As\ is less than high cross helicity \As.
In the intermediate range, the \As's uncertainties overlap with the low and high cross helicity values.
\cref{tbl:data} provides the fit parameters and their uncertainties, we summarize the behavior in the low and high \xhel\ ranges by their weighted means and include the intermediate range as a third group because it is not clearly a part of the other two.
We consider the low \xhel\ range of \As\ to be $\xhel\ \leq 0.51$ and the weighted mean of \As\ in this range is \As[3.87 \pm 0.04].
The saturation abundance increases by 0.14 percentage points to \As[4.01 \pm 0.06] for $0.51 < \xhel \leq 0.65$.
In the high \xhel\ regime with $0.65 < \xhel$, \As[4.13 \pm 0.01], a 0.12 percentage point increase over the intermediate regime and a 0.28 percentage point increase over \As\ in the low \xhel\ range.

\plotNdensVsw

\subsection{The Distribution of \vsw\ Observed near \au[1] During Solar Minima \label{sec:analysis:vsw-hist}}

To contextualize \vs\ in relationship to the peak of the slow wind (\vslow) during solar minima in \cref{fig:vsw-hist}, \cref{fig:ndens-vsw} plots the hydrogen (solid green) and helium (dash-dotted orange) number densities as a function of \vsw.
Here, we have created 2D column normalized histograms as in \cref{fig:ahe-vsw,fig:xhel-vsw}, selected data within 60\% of each column's maximum, and calculated the mean and standard deviation of these observations.
\cref{fig:ndens-vsw} plots the means with semi-transparent bands indicating each species' standard deviations.
The \n[\He] plot highlights four key speed ranges, which are given in \cref{tbl:speeds}, in contrasting colors.
Purple indicates the values of \n[\He] falling within the widths of the Gaussians fit to the slow and fast solar wind peaks characteristic of solar minima in \cref{fig:vsw-hist}.
Blue indicates the range of \n[\He] falling within the range of \vs\ as derived in \cref{fig:sat-fits} across \xhel\ and accounting for uncertainties.
The dark orange region labeled \vn\ indicates standard deviation of the peak of \n[\He] when the orange line is recalculated for all 15 \xhel\ quantiles.
The gray region indicates \vi, the intersection of the Gaussians fit to the slow and fast wind peak during solar minima.

\SpeedTable
We observe that \n[\Hy] decreases with increasing \vsw, as expected \citep[Figure 3]{LeChat2012}, from \n[\Hy][9.9 \pm 2.6] to an asymptotic value of \n[\Hy][2.6 \pm 0.3].
In contrast, \n[\He] reaches a local maximum of \n[\He][0.2 \pm 0.08] at \vn[415] for all the data, which is faster than the minimum \vs\ across \xhel.
Repeating this calculation for the 15 \xhel\ quantiles yields \vn[409 \pm 15].
For speeds \vsw*[\vn][<], \n[\He] increases monotonically from \n[\He][0.12 \pm 0.07] to its maximum.
For speeds \vsw*[\vn][>], \n[\He] decreases monotonically like \n[\Hy] to an asymptotic value of \n[\He][0.12 \pm 0.02].
As \vn[409 \pm 15] and \vn[415] are statistically indistinguishable, we will use the \vn[409 \pm 15] value to account for variability across \xhel.
The typical mean of the observed hydrogen densities across this range of speeds weighted by their standard deviations is \n[\Hy][5.3 \pm 0.43], where this uncertainty is the standard error of the mean.

\plotVswSolMin
\cref{fig:vsw:SolMin} plots the solar wind speed distribution during solar minima from \cref{fig:vsw-hist} and highlights the regions corresponding to each speed in \cref{tbl:speeds}: \vslow, \vn, \vs, \vi, and \vfast.
The colors match \cref{fig:ndens-vsw}.
The slow and fast wind peaks \vslow\ and \vfast\ are in purple.
The orange region is \vn.
The blue region corresponds to \vs.
The intersection between Gaussians fit to the slow and fast wind peak is \vi\ and indicated in green.
The \vfast\ peak is at the expected high speeds.
\vslow\ marks the peak of the slow wind.
\vs\ is just faster than \vslow\ and \vn\ spans the fast range of \vslow\ and slow range of \vs.
The slowest portion of \vi\ is just faster than the fastest portion of \vs.
The visual discrepancy between the speeds indicated here and the speeds indicated in \cref{fig:ndens-vsw} is related to the bin resolution in each.
That \vslow\ and \vs\ are adjacent and \vn\ spans these two speeds suggests that an unaccounted for variable may be relevant.

\subsection{Defining the $\left(\xhel,\ahe\right)$-Plane \label{sec:analysis:classification}}

\plotXhelAheVswContourArray*
\cref{fig:xhel-ahe-vsw:contours} is three contour plots of \func{\vsw}{\xhel,\ahe}.
The panels show 
\begin{inparaenum}[(a)]
\item mean \vsw, 
\item the 10\% \vsw\ quantile, and
\item the 90\% \vsw\ quantile.
\end{inparaenum}
Panels (b) and (c) use the same color scale, while Panel (a) uses a narrower speed range for the color scale.
Labeled contours are relevant speeds from \cref{fig:vsw:SolMin,tbl:speeds} derived in Panel (a) for mean \vsw; units are in \kms.
The color for each contour depends on the panel and is chosen for contrast.
In Panel (a), pink lines with a dash and three dots are the mean and upper bound of \vslow; black lines with a dash and five dots are \vi\ and its lower bound; and solid blue lines are the upper and lower bound of \vs.
Contours are smoothed with a 1$\sigma$ Gaussian filter for visual clarity.

In Panel (a), we observe two distinct regions of the mean \vsw\ plane with a third between them.
Solar wind with \vsw*[\vs][<] (blue to green regions) has contours that range from constant \ahe, irrespective of \xhel, or decreasing with \ahe\ as \xhel\ increases.
Excluding \ahe[8.5][>] with \xhel[0.6][\lesssim] (which is likely transients), speeds with \vsw[460][\geq] (orange to red) is linked to \xhel[0.73][\gtrsim] and for which the range of \ahe\ decreases with increasing \xhel.
Between 440 and \kms[460] (light to medium orange) lies a region of the plane where the contours follow neither trend.
This range includes \vi[450], the lower bound of the speed at which fast and slow wind solar wind peaks intersect.
Panels (b) and (c) display similar trends, but for different ranges of speeds.

Panel (b) plots the 10\% quantile of \vsw.
As stated above, the labeled speed contours correspond are derived from mean \vsw\ in Panel (a).
The entire range of speeds plotted is \kms[<390].
In fact, the region of the mean \vi[484] in highly Alfvénic wind shows that the 10\% quantile covers the speed range \vsw*[360] to \kms[390].

Panel (c) plots the 90\% quantile of \vsw.
Again, the labeled contours are derived from mean \vsw\ in Panel (a).
Here, the majority of the plane has speeds \vsw[400][>], covering the range of values typical of the \emph{ad hoc} speeds used to separate out or select for slow or fast wind.
In fact, the region of the plane where mean \vsw\ corresponds to the slow wind peak has speeds that range between \vsw*[420] and \kms[510].

\section{Discussion \label{sec:disc}} 

\subsection{Summary of Observations \label{sec:disc:sum}}

We have analyzed solar wind observations from the Wind/SWE Faraday cups covering solar cycles 23, 24, and the ascending phase of solar cycle 25.
From these observations, we have derived the solar wind speed (\vsw), helium abundance (\ahe), and normalized cross helicity (\xhel) on a 1-hour time scale.
We exclude speeds \vsw[300][<] because solar wind with those speeds has a vanishingly small \ahe\ that likely correspond to the very slow solar wind, which may uniquely originate from the heliospheric plasma sheet \citep[HPS,][]{Sanchez-Diaz2016a}.

The core of our analysis quantifies the change in the gradient of \ahe\ as a function of \vsw\ and how these changes in gradient vary with \xhel.
We identify the speed at which this gradient changes as the saturation speed (\vs), the corresponding abundance as the saturation abundance (\As), and the corresponding coordinate in the $\left(\vsw,\ahe\right)$-plane as the saturation point \satpoint.

For speeds \vsw*[\vs][<], we made observations about the minimum solar wind speed observed for \He\ and the gradients of \ahe\ as a function of \vsw.
First, \cref{fig:sat-fits:scaled} shows that the gradients of \ahe\ for \vsw*[\vs][<] are identical. 
This suggests that the process responsible for \ahe's strong gradient with \vsw\ in this regime is independent of \xhel\ and therefore independent of wave-particle interactions associated with Alfvén waves.
Second, \cref{fig:ahe-vsw} shows that the minimum \vsw\ for which there is a non-vanishing \ahe\ is \vv[302 \pm 4].
\cref{fig:sat-fits} further shows that this value may decrease with increasing \xhel\ by 10 to \kms[25], which is at most the same uncertainty found by \citet{Kasper:Ahe} when only considering \vsw*[\vs][<].
As \cref{fig:sat-fits:scaled} shows that the gradients of \ahe\ for \vsw*[\vs][<] are identical, further investigation is required to determine if this variation of \vv\ with \xhel\ is meaningful.

For speeds \vsw*[\vs][>], we similarly made observations about the gradients of \ahe\ as a function of \vsw\ and the extreme values of \ahe\ in the fastest wind.
\cref{fig:sat-fits,fig:sat-fits:scaled} show that the gradients of \ahe\ as a function of \vsw\ increase with decreasing \xhel.
We also observe that \ahe\ with low \xhel\ for these speeds exceeds \ahe\ with high \xhel\ at \vsw[525][\approx].
Qualitatively, \ahe\ with low $\xhel\ < 0.5$ becomes larger than \ahe\ with high $\xhel\ > 0.75$.
A detailed study of what drives this change in gradient at speeds \vsw[\vs][>] across \xhel\ is the subject of future work.

\cref{fig:dual_sat-xhel} (a) plots this saturation speed \vs\ in 15 \xhel\ quantiles.
The horizontal pink dotted line and the surrounding pink region is \vs\ calculated in \cref{fig:ahe-vsw} and its $1\sigma$ fit uncertainty.
The blue bars indicate \vs\ characteristic of low $\xhel\ \leq 0.77$,  intermediate $0.77 < \xhel\ \leq 0.91$, and high $0.91 < \xhel\ \leq 1$ normalized cross helicity.
Across these range, \vsw\ drops with increasing \xhel\ from \kms[430 \pm 1] to \kms[420 \pm 2] and then \kms[410 \pm 2].

\cref{fig:dual_sat-xhel} (b) relates the observed decrease in \vs\ with increasing \xhel\ to changes in \ahe\ at \vsw*[\vs].
As with \vs, we have identified three intervals in \xhel\ with different \As.
They are low $\xhel\ \leq 0.51$,  intermediate $0.51 < \xhel\ \leq 0.65$, and high $0.65 < \xhel\ \leq 1$ normalized cross helicities.
The corresponding \As\ are $3.87 \pm 0.04 \%$, $4.01 \pm 0.06 \%$, and $4.13 \pm 0.01 \%$.
Although the change in \emph{percentage points} is small, the percentage change in \As\ across these intervals in comparison to \As[4.19] calculated in \cref{fig:ahe-vsw} is larger than the percentage change in \vs\ across its \xhel\ intervals when referenced to the \cref{fig:ahe-vsw} value.
A detailed study of helium and hydrogen temperatures is necessary to determine if such an interpretation is supported by models suggesting that energy is taken from helium in the low solar atmosphere to accelerate hydrogen \citep[e.g.][]{Hansteen1997}.

The discrepancies between the \emph{Low}, \emph{Mid}, and \emph{High} \xhel\ ranges determined from \vs\ and \As\ are not unexpected.
We have crudely characterized the saturation point \satpoint\ with the intersection of two lines.
There is no reason to assume that a smoother functional form would not smooth out the trend of saturation points highlighted in \cref{fig:sat-fits}'s insert and lead to smoother curves in \cref{fig:dual_sat-xhel}.
However, smoother curves would be unlikely to remove the overall trends.

We also cannot rule out a solar activity component to these trends.
\citet[Figure 3]{DAmicis2021} shows that low values of \xhel\ are more common during solar minima than solar maxima.
\ahe\ is known to be strongly correlated with solar activity \citep{Wind:SWE:bimax,Alterman2019,Alterman2021,ACE:SWICS:AUX,Yogesh:Ahe}.
The relative occurrence rate of various solar sources along with their latitudinal stratification vary with solar activity \citep{McIntosh2015a,Hewins2020,Wang2002,Tlatov2014,Hathaway2015}.
As such, an underlying variability with solar activity does not rule out the overall trends we observe nor does it change our interpretation. 
It simply implies a time dependence of the source region occurrence rates and the impact of this time dependence on the saturation point \satpoint\ should be investigated in future work.
\sect{disc:asw} discusses the implications of these observations for the origin of the Alfvénic slow wind.

To contextualize the reported speeds, \cref{fig:ndens-vsw} breaks \ahe\ down into its component parts--\n[\Hy] and \n[\He]--and plots both as a function of \vsw.
\n[\Hy] monotonically decreases from \cc[9.9\pm2.6] to \cc[2.6 \pm 0.3] over the range of speeds 300 to \kms[800].
In contrast, \n[\He] monotonically increases from \cc[0.12 \pm 0.07] at \kms[300] to a maximum of \cc[0.2 \pm 0.08] at \vn[409 \pm 15].
Repeating this plot of \n[\He] for each \xhel\ quantile yields an average \vn[409 \pm 15], which spans the slow range of \vs\ and fast range of \vslow.
At speeds \vsw*[\vn][>], \n[\He] monotonically decreases to an asymptotic value of \cc[0.12\pm0.02].
\cref{sec:disc:a:ssw} discusses how this inflection in \n[\He] as a function of \vsw\ may point to the importance of \He\ in solar wind acceleration.

The ordering of \vslow, \vn, \vs, and \vi\ is unexpected and it would be unsurprising if one or more additional hidden variables are relevant to understanding the solar wind for \vsw*[\vs][<], especially the traditional transition between slow and fast wind as identified by the distinct peaks in a \vsw\ histogram for observations from solar minima.
\cref{fig:xhel-ahe-vsw:contours} plots \func{\vsw}{\xhel,\ahe} using (a) mean \vsw, (b) the 10\% \vsw\ quantile, and (c) the 90\% \vsw\ quantile with key contours of mean \vsw\ highlighted.
We observe that \vslow, \vs, and \vi\ define different domains of \vsw\ in the $\left(\xhel,\ahe\right)$-plane.
In general, for mean $\vsw*[\max\left(\vs\right)][<] = \kms[439]$, the \ahe\ to which a given speed contour corresponds decreases with increasing \xhel.
The strength of this gradient increases with increasing \vsw.
While the \vsw\ indicated in color changes with the different averaging schemes in each panel, the overall trends are the same.
For mean \vsw[460][>] in Panel (a) 
we observe that contours of constant \vsw\ are limited to \xhel[0.73][\gtrsim] and the range of \ahe\ for these contours of constant \vsw\ decreases with increasing \xhel.
A similar trend of maximum plotted \vsw\ being limited to a high range of \xhel\ and \ahe[3][\gtrsim] is also observed in Panels (b) and (c).
Building on Sections \ref{sec:disc:asw} and \ref{sec:disc:a:ssw}, \sect{disc:cat} argues that the $\left(\xhel,\ahe\right)$-plane can \emph{statistically} categorize the solar wind speed as originating from magnetically closed or magnetically open regions.

\cref{tbl:speeds} summarizes the speeds reported in this paper: \vslow, \vn, \vs, \vi, and \vfast.
\cref{fig:vsw:SolMin} plots the solar wind speed distribution during solar minima from \cref{fig:vsw-hist} and highlights the regions corresponding to each speed.
Building on prior Discussion sections, \sect{disc:vsw-SolMin} suggest that different portions of the bimodal distribution of solar wind observations observed during solar minima are dominated by different physical mechanisms.

\subsection{The Origin of the Alfvénic Slow Wind \label{sec:disc:asw}}

The two state solar wind emerging from different source regions with distinct magnetic topologies is a well-supported interpretation of the solar wind \citep{Schwenn2006,LR:CH}.
However, classifying the solar wind as fast or slow based on its speed is known to be inaccurate.
For example, the Alfvénic slow wind is an emerging \rd{3} class of solar wind with speeds that are most similar to the slow wind but with other kinetic, chemical, charge-state, and Alfvénic properties that are typical of fast solar wind \citep{DAmicis2021a,DAmicis2021,DAmicis2018,Damicis2016,DAmicis2015}.
Although recent theories and observations would attribute a key role in it generation to the strong divergence of open magnetic field lines \citep[e.g.][]{Panasenco2013,Panasenco2019,Panasenco2020}, the source of the Alfvénic slow wind is yet-to-be-determined.

Under the two-state fast/slow paradigm, the transition between fast and slow wind happens between 400 and \kms[600].
Defining arbitrary time intervals that correspond to solar minima and \emph{ad hoc} thresholds below and above which to consider slow and fast wind, \citet{Wind:SWE:Wk} fit the peaks of fast and slow solar wind during solar minima in \cref{fig:vsw-hist} with Gaussians and identify a fast/slow transition under the two-state paradigm at  \vi[484 \pm 34] based on the intersection of these Gaussians.
\cref{fig:ahe-vsw} identifies a saturation speed \vs[433 \pm 4] based on the change in gradient of \ahe\ (\grad[\ahe][\vsw]) across this speed.
This \vi\ is \kms[54] faster than \vs[430 \pm 1] in non-Alfvénic (low \xhel) wind and \kms[74] faster than \vs[410 \pm 2] in Alfvénic (high \xhel) wind.
Because \vs\ corresponds to \As\ and \As\ is the value of \ahe\ characteristic of fast solar wind, this means that there is a \kms[54] to \kms[74] interval of \vsw\ for which the two state paradigm's slow wind has \ahe\ characteristic of fast solar wind.
Because \vs\ decreases with increasing \xhel, the size of this interval increases with \xhel.
This means that the more Alfvénic the solar wind, the more likely that that an observation classified as slow under the two-state paradigm based on \vsw\ may have properties more similar to typical fast wind.
In other words, there is a significant fraction of the solar wind with slow speeds and other fast wind-like properties that is incompatible with the fast/slow paradigm.
This is the Alfvénic slow wind.

Considering how \satpoint\ evolves with \xhel\ along with the gradients of \ahe\ with \vsw\ above and below saturation addresses this contradiction. Recall the following observations from \sect{analysis:ahe-xhel}.
\cref{fig:sat-fits,fig:sat-fits:scaled} show that the gradient of \ahe\ as a function of \vsw\ does not vary with \xhel\ for speeds \vsw*[\vs][<].
Above \vs, the gradient drops and the magnitude of this change in \grad[\ahe][\vsw] increases with increasing \xhel.
The insert in \cref{fig:sat-fits} shows an anti-correlation between \vs\ and \As.
\cref{fig:dual_sat-xhel} demonstrates that this anti-correlation is a function of \xhel:
\As\ is smaller and \vs\ is larger in low \xhel;
\As\ is larger and \vs\ is smaller in high \xhel.

We infer that solar wind from below the saturation point with \vsw*[\vs][<] and \ahe*[\As][<] is accelerated in magnetically closed source regions based on the following observations:
\begin{compactenum}
\item \ahe\ is highly variable for speeds \vsw*[\vs][<], as typical for solar wind from closed loops and other (typically) equatorial sources.
\item The gradient of \ahe\ with \vsw\ is independent of \xhel\ for speeds \vsw*[\vs][<], indicating that \grad[\ahe][\vsw] for \vsw*[\vs][<] is independent of wave-particle processes related to the presence of Alfvén waves that are progressively more common as \xhel\ increases.
\end{compactenum}
Based on the following observations, we infer that solar wind from above the saturation point is accelerated in magnetically open source regions.
\begin{compactenum}
\item The gradient of \ahe\ increases with decreasing \xhel\ for \vsw*[\vs][>].
\item The discrepancy between \ahe's gradient with \vsw\ for speeds \vsw*[\vs][<] and \vsw*[\vs][>] increases with increasing \xhel.
In other words, the upper limit on \ahe\ becomes progressively tighter or more stringent as \xhel\ increases.
\end{compactenum}
Together, these observations suggests that \grad[\ahe][\vsw] for \vsw*[\vs][>] depends on Alfvénic content of the solar wind and the related wave-particle processes that are increasingly more common as \xhel\ increases.
Combining these inferences about the relationship between the saturation point and the magnetic topology of source regions, the maximum speed of solar wind from magnetically closed sources is larger than the minimum speed of solar wind from magnetically open sources.
The natural conclusion from these inferences is that Alfvénic slow wind is likely solar wind accelerated at magnetically open sources, which results in kinetic, chemical, and charge-state properties typically associated with fast wind and typical of CHs.
The speed of such Alfvénic slow wind, which can have $\vsw < $ non-Alfvénic \vs\ from magnetically closed sources, is simply a result of the possible ranges of speeds compatible with a magnetically open topology.
Such an interpretation is consistent with \cref{fig:xhel-ahe-vsw:contours}.

\subsection{Solar Wind Acceleration \label{sec:disc:a:ssw}}

An increase in momentum below the solar wind's sonic point increases the mass flux and leads to a decreased asymptotic solar wind speed at \au[1] (\vau) because it decrease the energy per particle \citep{Leer1980,Holzer1981,Holzer1980a,Holzer1981}.
This effect is larger than the increase in \vau\ due to energy addition in the subsonic region.
A natural consequence of this mass flux effect on \vsw\ is that the solar wind's mass density decreases with increasing speed.

\citet{Lie-Svendsen2001,Lie-Svendsen2002} show that the solar wind mass flux and speed are driven by the coronal heat flux into the chromosphere/transition region, which is a function of the coronal magnetic topology.
In open field regions, Alfvén waves provide sufficient energy to lift the solar plasma and accelerate it into the solar wind \citep{Leer1980a}.
As such, \He\ is a minor species that simply provides an additional mass term in the solar wind mass flux, any excess non-thermal energy in \He\ (e.g. super-mass-proportional temperatures) remains with it, and \He\ behaves similarly to protons \citep{Endeve2005,Lie-Svendsen2003,Hansteen1997}.
In such regions, neither the \Hy\ nor \He\ flux in the solar wind is highly sensitive to the heat flux mediated coupling between the corona and deeper layers of the Sun.
In closed field regions, which dominate observations with speeds \vsw*[\vs][<], there is insufficient energy to simply lift the solar plasma and accelerate it into the solar wind \citep{Endeve2005}.
As such, collisional coupling between \He\ and \Hy\ transfer any excess non-thermal energy from \He\ to \Hy\ so that \Hy\ can escape, forming the solar wind \citep{Hansteen1997,Endeve2005}.
Only after \Hy\ has escaped the Sun and its density has dropped does the collisional coupling between \Hy\ and \He\ become sufficiently inefficient that \He\ retains sufficient energy to be accelerated into the solar wind \citep{Endeve2005}.
Because these closed regions only open intermittently and for variable lengths of time, the \ahe\ from them is highly variable \citep{Endeve2005}.
In these regions, the heat flux drives the amount of mass that enters the corona and therefore settles in these loops, making the helium abundance highly sensitive to both the closed field topology and the heat flux from the corona into the transition region \citep{Endeve2005}.

\cref{fig:ndens-vsw} plots the hydrogen (\n[\Hy]) and helium (\n[\He]) number densities as a function of \vsw.
\n[\Hy] decreases monotonically with \vsw\ 
\n[\He] reaches a local maximum of \n[\He][0.2] at \vn[415], which varies as \vn[409] across \xhel.
At the extreme speeds, \n[\He][0.12] with the standard deviations of \cc[0.07] at \kms[300] and \cc[0.02] at \kms[800].
Like \n[\Hy], \n[\He][0.12] is an asymptotic value at the fastest speeds.

\cref{fig:ndens-vsw} shows that the slowest \vs\ in the most Alfvénic wind (highest \xhel), which is most characteristic of solar wind accelerated in magnetically open sources, is \kms[8] faster than the fastest \vslow\ accounting for the slow wind peak's width.
This difference is $2\times$ to $3\times$ smaller than the 17 to \kms[23] difference between \vs\ in the low \xhel\ and high \xhel\ ranges.
In contrast, the range of \vn\ (calculated as an average over \xhel\ quantiles) overlaps \vslow\ for \kms[13] and the majority of \vs's range.
\vn\ is the speed at which \n[\He] peaks as a function of \vsw.
From this, we infer that the signature of magnetically open sources becomes significant at speeds just faster than those characteristic of the slow wind and slower than \vfast.
Across this transition between magnetically closed and open sources, the gradient of \n[\He] with \vsw\ changes.
The hydrogen number density typical of these speeds at which \n[\He] has an inflection point is \n[\Hy][5.3 \pm 0.43][>].
\cref{tbl:speeds} summarize these speeds.

Together, 
\begin{inparaenum}[(1)]
\item the change in gradient of \n[\He] with \vsw\ when \n[\Hy] has dropped to approximately \cc[5.3];  
\item that this change in gradient is intensive to changes in \xhel; 
\item the increased significance of solar wind from magnetically open sources at speeds \vsw*[\vslow][>]; 
\item the range of speeds over which this change in \n[\He]'s gradient with \vsw\ occurs; and
\item that \ahe\ is set below the sonic point at $r < \rc$
\end{inparaenum}
may suggest there is insufficient energy in the subsonic region for helium to be continuously accelerated into the solar wind with hydrogen at close field sources.
That 
\begin{inparaenum}[(1)]
\item \n[\He] decreases with increasing \vsw\ for speeds \vsw*[\vslow][>]; 
\item that \vs\ is a signature of the solar wind's transitions from magnetically closed to magnetically open sources at these speeds; and
\item that these speeds are faster than those characteristic of the slow wind peak
\end{inparaenum}
may suggest that there is sufficient energy in the subsonic region below open magnetic topologies that helium is collisionally decoupled from hydrogen and it retains sufficient energy to be accelerated into the solar wind \citep{Lie-Svendsen2003,Endeve2005}, simply providing, ``an [additional] effective momentum term'' \citep{Leer1979a}.
A detailed analysis of helium and hydrogen temperatures is necessary to determine if such an interpretation is supported by models suggesting that energy is taken from helium in the low solar atmosphere to accelerate hydrogen \citep[e.g.][]{Hansteen1997}.

This interpretation is consistent with \citepossessive{Alterman2019} work.
They argue that the long term delay in slow wind \ahe's response to changes in sunspot number (\SSN) is due to a physical process characteristic of slow wind source regions that depletes \ahe\ from its fast wind values.
One possibility they propose is gravitational settling in long-lived magnetic loops, which are characteristic of such close field regions \citep{Lenz1998,Uzzo2003,Uzzo2004,Rivera:CME}.
Such loops are also related to the coupling of the heat flux into the chromosphere and transition region \citep{DePontieu2009,Hou2024,Hansteen1997}.
Under such an interpretation, the decrease \n[\He] with decreasing \vsw\ for speeds \vsw*[\vn][<] is due to gravitational settling depleting \n[\He] from the value implied by its fast wind trends when \He\ is not accelerated into the solar wind.
If this is the case, then the typical \ahe\ at heights below those where open and closed magnetic topologies become significant would have an average maximum value of \ahe[4.19 \pm 0.05], which is $49 \pm 2\%$ of photospheric \ahe[8.2], inferred from helioseismology \citep[and references therein]{Asplund2021}.

Although we cannot explain why this would be the case, the unexpected minimum \n[\He][0.12] at 300 and \kms[800] may indicate a minimum helium particle density is necessary for helium particles to be accelerated into the solar wind.
However, this seems unlikely. 
The minimum \n[\He] at \kms[300] is not an asymptotically small value and we cannot rule out instrument limitations that lead to us excluding smaller \n[\He] at speeds \vsw[300][<], but have been excluded from this study.

\subsection{Categorizing Solar Wind Observations in the $\left(\xhel,\ahe\right)$-Plane \label{sec:disc:cat}}

\citet{Viall:9Q} define 9 open questions in solar wind physics that represent major outstanding problems.
The first three focus on solar wind formation:
\begin{compactenum}
\item ``From where on the Sun does the solar wind originate?''
\item ``How is the solar wind released?''
\item ``How is the solar wind accelerated?''
\end{compactenum}
Answering these questions relates \emph{in situ} observations to
\begin{inparaenum}[(1)]
\item source regions
\item with distinct release mechanisms that
\item lead to distinct asymptotic solar wind speeds.
\end{inparaenum}

We have characterized a signature of the transition between magnetically closed and magnetically open sources observed in \emph{Wind} observations.
This signature combines helium abundances and cross helicity observations.
The helium abundance (\ahe) is a signature of physics set below the sonic critical point ($r < \rc$), likely in the chromosphere and/or transition region \citep{LR:FIP,Schwadron1999,Laming2004,Rakowski2012,Endeve2005,Lie-Svendsen2003}.
The normalized cross helicity (\xhel) measures the Alfvénicity of a given \emph{in situ} observation on a relevant timescale.
We chose 1hr, a typical Alfvénic scale in the MHD domain.
The cross helicity is necessarily set above the sonic critical point ($r > \rc$), only decays above the Alfvén point ($r > \rA$), and quantifies the significance of Alfvén waves as indicated by high \Bb-\Bv\ correlations in a given \emph{in situ} solar wind observation.

\cref{fig:xhel-ahe-vsw:contours} (a) plots contours of constant mean \vsw\ as a function of \xhel\ and \ahe.
The labeled contours identify key speeds from \cref{tbl:speeds}.
The plane is separated into three regions.
The region corresponding to \vsw*[\min\left(\vs\right)][\leq] covers all possible \xhel\ and a reduced range of \ahe\ as \xhel\ increases.
The region corresponding to \vsw[460][\geq] covers \ahe\ that are larger than the \vsw*[\max\left(\vslow\right)] region and limited to \xhel[0.73][\gtrsim].
Panel (b) follow the same format as Panel (a), plotting the 10\% \vsw\ quantile with an increased range of \vsw\ indicated in by the color.
Panel (c) does the same for the 90\% \vsw\ quantile.
Comparing these panels, we notice that the 10\% level in the highly Alfvénic region is smaller than the 90\% level in the low \ahe\ region of the plane.
Due to a combination of several processes that are set below the sonic point, solar wind originating in closed or intermittently open field regions typically has \ahe\ that is smaller than \ahe\ in solar wind from open field regions.
At the same time, the dynamical processes that lead to the intermittent opening of the magnetic field in these regions likely leads to a wide range of wave modes while continuously open fields in regions with predominantly radial magnetic fields are more likely to preferentially carry Alfvén waves.
This suggests that solar wind
from close field regions has \ahe\ of at most 4.25\% and this upper bound on \ahe\ in close field regions decreases as the process through which the source region field becomes intermittently open leads to higher Alfvénicity (\xhel).
In contrast, solar wind from open field regions has \xhel[0.73][\gtrsim], carries \ahe[2.6][\gtrsim], and the minimum \ahe\ of such solar wind increases as \xhel\ decreases.
Based on \cref{fig:xhel-ahe-vsw:contours}, the \emph{statistical} upper bound of solar wind that may be considered from closed source regions is \vsw[439].
In other words, the variation of $\min\ahe$ in solar wind born in open magnetic field regions and $\max\ahe$ born in closed magnetic filed regions reflects the energy partition between \Hy\ and \He\ governed by the details of the solar wind's acceleration in these regions.

\plotXhelAheVswContoursCategorization

Interpreting the range of speeds $\kms[439] <$ mean $\vsw[460][\leq]$ is less straightforward.
Comparing Panels (b) and (c) shows that the 10\% and 90\% speed quintiles cover speeds from 360 to \kms[600].
We suspect that this \kms[21] wide mean \vsw\ interval likely corresponds to the region of the $\left(\xhel,\ahe\right)$-plane where solar wind emanating from open and closed field regions mixes and the frequency at which the spacecraft samples solar wind originating in the two classes of source regions is significant.
Further investigation of the detailed behavior of the solar wind abundances and temperature ratios is necessary to confirm such a suspicion.

\cref{fig:xhel-ahe-vsw:contours:categorization} is a semi-quantitative cartoon summaring of the preceding two paragraphs.
Here, we have plotted two contours of constant \vsw\ in the $\left(\xhel,\ahe\right)$-plane at \vsw*[425] and \kms[460] that separate the plane into three regions, which are labeled on the plot.
The \vsw[425] contour is within the range of \vs\ and chosen by eye so that \func{\ahe}{\xhel[0]}[4.19 = \As][\approx].
The gray region indicates \vs*[407] to \kms[439] and indicates the range over which we somewhat arbitrarily could have specified any contour.
This green region is where we consider the solar wind to have originated from closed field source regions on the Sun.
The \vsw[460] contour is chosen so that the contour on the right hand side of the plot is smooth and, excluding the \emph{Transients} region, bounded.
The region on the right side of the plot is where we consider solar wind from open field regions to have originated.
The \emph{Mixed Source} region is the region of the plane where solar wind from open and closed field regions is observe and the structure of the $\left(\xhel,\ahe\right)$-plane likely depends on the frequency with which solar wind from each source is sampled by a given spacecraft.
Because of this sample frequency consideration, we emphasize that the choice of \vsw[425] to identify solar wind from close field regions is somewhat arbitrary and within the range of \vs, which covers the gray region of the plane.
A manuscript \emph{in prep} publication shows that transients are the source of high speed solar wind in the top left corner, where the mean \vsw\ is the same as the open field region, and that when transients are removed the mean \vsw\ in this \emph{Transient} region is the same as the \emph{Mixed Source} region.

\subsection{The Distribution of Solar Wind Speeds at \au[1] During Solar Minima \label{sec:disc:vsw-SolMin}}

The solar wind's bimodal distribution is most prominent during solar minima.
Our categorization scheme in the $\left(\xhel,\ahe\right)$-plane and the key speeds derived in \cref{tbl:speeds} enable us to attribute different regions of the distribution to different physical processes.
The following summary highlights the mixing of source regions in the yellow region of \cref{fig:xhel-ahe-vsw:contours:categorization} and suggests a categorization for \vsw\ if observations of \ahe\ and \xhel\ are unavailable.
In the following list, \vsw\ refers to mean \vsw\ as presented in \cref{fig:vsw:SolMin,fig:xhel-ahe-vsw:contours} panel (a) unless otherwise specified.
\begin{compactenum}
\item Observations with \vslow[399][<] is likely dominated by solar wind originating in closed field regions.
\item Speeds \vslow[399] to \vs[407] is the slowest range of speeds where solar wind originating in closed and open field source regions are observed to mix when observed at \au[1].
\item In the speed range \vs[407] to \vs[439], the helium abundance from open field regions begins to become the dominant helium abundance signature at \au[1].
\item In the range of speeds \vs[439] to \vi[484], \au[1] observations of solar wind are dominantly from open field regions, but solar wind from closed field regions still compose a non-trivial fraction of the solar wind observations.
This \vsw\ is higher than the speed specified in \cref{fig:xhel-ahe-vsw:contours:categorization} because the $\left(\xhel,\ahe\right)$-plane better separates solar wind from open and closed source regions.
\item This slow wind ``contamination'' of the fast wind signatures decreases over the speed range \vi[484] to \vfast[564].
\item For speeds \vfast[564][\geq], the solar wind can be considered to originate exclusively in open field source regions.
\end{compactenum}
These speed ranges clarify why the threshold between fast and slow solar wind typically set between \vsw*[400][\sim] to \kms[\sim600].
We emphasize that the above are \emph{statistical} trends and that either 
\begin{inparaenum}[(a)]
\item the \vsw\ threshold between solar wind from magnetically closed and open regions must be chosen \emph{ad hoc} for individual events because the mixing of open and closed field solar wind happens over the large range of speeds or
\item the identification of solar wind originating with closed or open magnetic topologies at the Sun's surface should be identified in the $\left(\xhel,\ahe\right)$-plane.
\end{inparaenum}

\section{Conclusion \label{sec:conclusion}}

Long-standing \emph{in situ} observations confirmed by many spacecraft show that the solar wind can be considered as a two-state system and that the two states can be divided according to their speed into fast and slow wind.
The fast/slow distinction is typically treated as a consequence of the solar source region in which the solar wind was accelerated: fast wind is considered to have originated in regions like coronal holes with magnetically open topologies while slow wind is considered to have originated in regions with magnetically closed topologies like the pseudostreamer and spearatrix web, which have magnetic fields that are only intermittently open to the heliosphere.
As a consequence of the different magnetic topologies in these regions, slow and fast wind carry distinct helium abundances (\ahe), Alfvénicities (\xhel*), chemical makeup, and charge-state ratios.
The Alfvénic slow wind is an emerging \rd{3} class of solar wind that violates this two state paradigm: the majority of its properties are typical of solar wind accelerated in regions with magnetically open source regions, but its speed is typical of solar wind accelerated in regions with magnetically closed topologies.
In other words, \vsw\ is alone insufficient for mapping the solar wind to its source regions.
This work directly and statistically addresses the origin of the Alfvénic slow wind.

The solar wind's helium abundance (\ahe) is set below the sonic critical point (\rc) and reflects processes in the Sun's chromosphere and transition region.
The solar wind's Alfvénicity is set near the Alfvén critical surface (\rA), which is above \rc, and reflects the magnetic topology of the source region.
Using 28 years of Wind observations at 1 AU, we have performed a long-duration statistical analysis of the relationship between \ahe, \xhel\, and \vsw.
We make the following observations.
\begin{compactenum}
\item We define the saturation point \satpoint\ as the solar wind speed and abundance for which the gradient of \ahe\ as a function of \vsw\ changes.
This speed is \vs[433 \pm 4].
    \begin{compactenum}
    \item For speeds \vsw*[\vs][<], \ahe\ is highly variable.
    \item For speeds \vsw*[\vs][>], \ahe[4.19 \pm 0.05] and remains constant.
    \end{compactenum}
\item \As[4.19 \pm 0.05] is equivalent to a solar wind helium abundance that is $49 \pm 2 \%$ of its photospheric abundance.
\item Repeating the analysis of the saturation point across 15 quantiles in \xhel\, we show that the saturation speed \vs\ and saturation abundance \As\ are anti-correlated and this anti-correlation is a function of \xhel.
\item The slowest saturation speed \vs, which corresponds to the most Alfvénic solar wind (highest \xhel), is \kms[8] faster than the fastest speed typical of the slow solar wind peak \vslow\ during solar minima. 
\item To contextualize the saturation point with respect to the slow wind, we show that $\n[\He]\left(\vsw\right)$ has a local maximum at \vn[409 \pm 15] averaged across 15 \xhel\ quantiles.
The range of \vn\ covers the fastest \vslow\ speeds and the slowest observed \vs.
\item \citet{Wind:SWE:Wk} quantify the peaks of the bimodal solar wind distribution at \au[1] during solar minima as \vslow[355 \pm 44] and \vfast[622 \pm 58].
These distributions intersect at \kms[484 \pm 34], well within the \emph{ad hoc} range of values at which the slow and fast wind are often separated.
Summarizing these speeds and those identified in this paper, we show that there is a 52 to \kms[75] wide interval from approximately 407 to \kms[484] for which solar wind is often identified as slow when it may be accelerated in magnetically open source regions.
Separating solar wind in these speeds as open and closed is sensitive to the solar wind sampled.
\item Plotting \vsw\ as a function of \xhel\ and \ahe\ in \cref{fig:xhel-ahe-vsw:contours}, we observe two distinct populations. 
    \begin{compactenum}
        \item At high \xhel[0.73][\gtrsim], solar wind speeds with mean \vsw[440][>] have \ahe[2.6][\gtrsim].
        \item For low \ahe[4.19][\lesssim], \ahe\ decreases with \xhel\ without a predominately preferred \xhel\ and \vsw\ can take on the range of speeds \vsw[510][\lesssim].
    \end{compactenum}
\cref{fig:xhel-ahe-vsw:contours:categorization} is a semi-quantitative cartoon illustrating this.
\end{compactenum}

From the observations of the anti-correlation between \vs\ and \As\ that depends on \xhel\, we infer that the maximum speed of non-Alfvénic solar wind from closed field regions is faster than the minimum speed observed for Alfvénic solar wind from open field regions.
The natural consequence is that the Alfvénic slow wind is solar wind from magnetically open sources with speeds that fall into this overlapping speed interval.
We also show that, because \ahe\ is set below the sonic critical point (\rc) and \xhel\ is set near the Alfvén surface (\rA), categorizing the solar wind as a function of \xhel\ and \ahe\ 
statistically differentiates between solar wind from magnetically open and magnetically closed source regions.
However, unlike observations of elemental and charge-state composition, such a categorization scheme does not require data from an ion composition mass spectrometer.

Combining the speed ranges summarized in \cref{tbl:speeds}, the categorization scheme presented in \cref{fig:xhel-ahe-vsw:contours}, and our understanding of the relationship between the saturation point \satpoint\ and \xhel\ allows us to characterize the distribution of solar wind speeds observed during solar minima at \au[1].
\cref{fig:vsw:SolMin} identifies the speeds relevant for different regions of this distribution and provides a suggested categorization if observations of \ahe\ and \xhel\ are unavailable.
\sect{disc:vsw-SolMin} provides an itemized list that explains the different portions of this distribution and explains why the threshold between fast and slow solar wind typically set between \vsw*[400][\sim] to \kms[\sim600] in an \emph{ad hoc} fashion.

Drawing on coupled models of the chromosphere, transition region, corona, and solar wind \citep{Lie-Svendsen2001,Lie-Svendsen2002,Lie-Svendsen2003,Endeve2005}, we then draw the following conclusions.
\begin{compactenum}
\item In close field regions, there is insufficient energy below the sonic point to continuously accelerate the solar wind.
Rather excess energy in the helium (e.g.\ super-mass-proportional temperatures) is collisionally transferred to hydrogen, which increases the energy of the latter sufficiently for it to leave the sun and become the solar wind when a closed field region temporarily opens.
Once \n[\Hy] has decreased sufficiently in these regions, the collisional coupling between helium and hydrogen drops, and helium retains the energy necessary to enter the solar wind.
\item In open field regions, there is sufficient energy to accelerate the solar wind and it leaves the corona with a Parker-like acceleration.
In these open field regions, \He\ provides an effective momentum term in the relevant equations and is accelerated in a similar manner as \Hy\ \citep{Leer1979a}.
\end{compactenum}
If these coupled models are correct, this suggests that the coupling between the transition region and corona, which is mediated by the downward coronal heat flux, is essential for setting the helium abundance in the solar wind and the solar wind helium abundance reflects the energy partition between \Hy\ and \He\ in open and closed source regions \citep{Endeve2005,Lie-Svendsen2003}.

Unexpectedly, we observe that \n[\He][0.12] at both 300 and \kms[800].
One possible interpretation is that there is a minimum \n[\He] for helium to be accelerated into the solar wind.
This may be consistent with the work showing that there is also a minimum speed at which helium is accelerated into the solar wind \citep{Kasper:Ahe,Alterman2019}.
However, we have difficulty believing that the minimum helium density corresponds to the value observed at \kms[300] because we have selected data for which \vsw[300][>].
Rather, a detailed examination of solar wind with speeds \vsw[300][<], which may originate in the heliospheric plasma sheet \citep[HPS,][]{Sanchez-Diaz2016a}, is necessary to characterize such a minimum \n[\He].

Finally, we note that this work highlights the essential role played by utilizing multiple \emph{in situ} parameters to relate solar wind observations to their source regions and the limitations of simply categorizing the solar wind based on its speed.
As such, it is particularly important in the content of the Solar Orbiter mission \citep{SO:A,SO:B}, launched in 2020, aimed at exploring the inner heliosphere where fundamental plasma processing such as solar wind heating and acceleration take place.
One of the aims of the mission is to answer significant questions regarding the solar sources of the slow solar wind, both Alfvénic and non-Alfvénic, and its evolution, which requires detailed characterization of the \emph{in situ} plasma.
The Solar Wind Analyzer suite on board Solar Orbiter \citep{SO:SWA} provides such observations.
This suite includes the Heavy Ion Sensor (HIS), which provides the first heavy ion observations  in the inner heliosphere within the orbit of Mercury \citep{SO:HIS}.

\begin{acknowledgments}
The authors are grateful to Stefano Livi, Mihailo M. Martinović, Yeimy J. Rivera, Kristopher G. Klein, Bennet A. Maruca, and Michael L. Stevens for valuable discussions.
They also thank the referee for their useful and insightful comments.
Wind Faraday cup data are obtained from CDAWeb and the authors gratefully acknowledge Justin C. Kasper for the development and Michael L. Stevens for the delivery of this data product.
B.L.A. is funded by NASA contract NNG10EK25C and grants 80NSSC22K0645 (LWS/TM), 80NSSC22K1011 (LWS), and 80NSSC20K1844.
The activity related to Solar Orbiter of R.D. is currently funded under Italian Space Agency (ASI) grant 2018-30-HH.1-2022.

\end{acknowledgments}

\software{
IPython \citep{Perez2007}, 
Jupyter \citep{Kluyver2016,Kluyver2016a}, 
Matplotlib \citep{Hunter2007}, 
Numpy \citep{Harris2020,VanderWalt2011}, 
SciPy \citep{Jones2001,scipy},
Pandas \citep{Mckinney2010,McKinney2011,Mckinney2013}, 
Python \citep{Millman2011,Oliphant2007},
Mathematica \citep{Mathematica:14.0}
} 
\bibliography{Zotero.bib}{}
\bibliographystyle{aasjournal}

\end{document}